\documentclass[12pt,preprint]{aastex}

\usepackage{amsmath}

\newcommand{\deV}{deVaucouleurs}
\newcommand{\sns}{SDSS-II Supernova Survey}

\newcommand{\nsnetot}{34}

\newcommand{\nsnecf}{9}
\newcommand{\cfsnur}{0.37}
\newcommand{\cfsnurhi}{0.17}
\newcommand{\cfsnurlo}{0.12}
\newcommand{\cfsnurhisyst}{0.01}
\newcommand{\cfsnurlosyst}{0.01}

\newcommand{\cfsnurearly}{0.31}
\newcommand{\cfsnurhiearly}{0.18}
\newcommand{\cfsnurloearly}{0.12}
\newcommand{\cfsnurhisystearly}{0.01}
\newcommand{\cfsnurlosystearly}{0.01}

\newcommand{\cfsnub}{0.46}
\newcommand{\cfsnubhi}{0.21}
\newcommand{\cfsnublo}{0.15}
\newcommand{\cfsnubhisyst}{0.01}
\newcommand{\cfsnublosyst}{0.01}

\newcommand{\cfsnum}{0.123}
\newcommand{\cfsnumhi}{0.056}
\newcommand{\cfsnumlo}{0.040}
\newcommand{\cfsnumhisyst}{0.004}
\newcommand{\cfsnumlosyst}{0.003}

\newcommand{\bcgeplt}{4.85 \times 10^{13}}
\newcommand{\nsnebcg}{27}

\newcommand{\nsnebcgnobias}{25}

\newcommand{\bcgsnur}{0.55}
\newcommand{\bcgsnurhi}{0.13}
\newcommand{\bcgsnurlo}{0.11}
\newcommand{\bcgsnurhisyst}{0.02}
\newcommand{\bcgsnurlosyst}{0.01}

\newcommand{\bcgsnurearly}{0.49}
\newcommand{\bcgsnurhiearly}{0.15}
\newcommand{\bcgsnurloearly}{0.11}
\newcommand{\bcgsnurhisystearly}{0.02}
\newcommand{\bcgsnurlosystearly}{0.01}

\newcommand{\bcgsnub}{0.68}
\newcommand{\bcgsnubhi}{0.17}
\newcommand{\bcgsnublo}{0.14}
\newcommand{\bcgsnubhisyst}{0.02}
\newcommand{\bcgsnublosyst}{0.02}

\newcommand{\bcgsnum}{0.18}
\newcommand{\bcgsnumhi}{0.044}
\newcommand{\bcgsnumlo}{0.036}
\newcommand{\bcgsnumhisyst}{0.006}
\newcommand{\bcgsnumlosyst}{0.004}

\newcommand{\cfsnurbcg}{2.04}
\newcommand{\cfsnurhibcg}{1.99}
\newcommand{\cfsnurlobcg}{1.11}
\newcommand{\cfsnurhisystbcg}{0.07}
\newcommand{\cfsnurlosystbcg}{0.04}

\newcommand{\bcgsnurbcg}{0.36}
\newcommand{\bcgsnurhibcg}{0.84}
\newcommand{\bcgsnurlobcg}{0.30}
\newcommand{\bcgsnurhisystbcg}{0.01}
\newcommand{\bcgsnurlosystbcg}{0.01}

\newcommand{\cfrat}{1.94}
\newcommand{\cfrathi}{1.31}
\newcommand{\cfratlo}{0.91}
\newcommand{\cfratsysthi}{0.043}
\newcommand{\cfratsystlo}{0.015}

\newcommand{\bcgrat}{3.02}
\newcommand{\bcgrathi}{1.31}
\newcommand{\bcgratlo}{1.03}
\newcommand{\bcgratsysthi}{0.062}
\newcommand{\bcgratsystlo}{0.048}

\begin{document}

\title{A Measurement of the Rate of Type Ia Supernovae in Galaxy Clusters 
from the SDSS-II Supernova Survey
}

\date{\today}

\email{bdilday@physics.rutgers.edu}

\author{
Benjamin~Dilday,\altaffilmark{1,2,3}
Bruce~Bassett,\altaffilmark{4,5}
Andrew~Becker,\altaffilmark{6}
Ralf~Bender,\altaffilmark{7,8}
Francisco~Castander,\altaffilmark{9}
David~Cinabro,\altaffilmark{10}
Joshua~A.~Frieman,\altaffilmark{11,12}
Llu\'{\i}s~Galbany,\altaffilmark{13}
Peter~Garnavich,\altaffilmark{14}
Ariel~Goobar,\altaffilmark{15,16}
Ulrich~Hopp,\altaffilmark{7,8}
Yutaka~Ihara,\altaffilmark{17}
Saurabh~W.~Jha,\altaffilmark{1}
Richard~Kessler,\altaffilmark{3,11}
Hubert~Lampeitl,\altaffilmark{18}
John~Marriner,\altaffilmark{12}
Ramon~Miquel,\altaffilmark{13,19}
Mercedes~Moll\'a,\altaffilmark{20}
Robert~C.~Nichol,\altaffilmark{18}
Jakob~Nordin,\altaffilmark{16}
Adam~G.~Riess,\altaffilmark{21,22}
Masao~Sako,\altaffilmark{23}
Donald~P.~Schneider,\altaffilmark{24}
Mathew~Smith,\altaffilmark{4,18}
Jesper~Sollerman,\altaffilmark{15,25}
J.~Craig~Wheeler,\altaffilmark{26}
Linda~\"{O}stman,\altaffilmark{16}
Dmitry~Bizyaev,\altaffilmark{27}
Howard~Brewington,\altaffilmark{27}
Elena~Malanushenko,\altaffilmark{27}
Viktor~Malanushenko,\altaffilmark{27}
Dan~Oravetz,\altaffilmark{27}
Kaike~Pan,\altaffilmark{27}
Audrey~Simmons,\altaffilmark{27}
and Stephanie~Snedden\altaffilmark{27}
}
\altaffiltext{1}{
Dept. of Physics and Astronomy,Rutgers, the State University of New Jersey,136 Frelinghuysen Rd., Piscataway, NJ 08854.
}
\altaffiltext{2}{
Department of Physics, University of Chicago, Chicago, IL 60637.
}
\altaffiltext{3}{
Kavli Institute for Cosmological Physics, The University of Chicago, 5640 South Ellis Avenue Chicago, IL 60637. 
}
\altaffiltext{4}{
Department of Mathematics and Applied Mathematics, University of Cape Town, Rondebosch 7701, South Africa. 
}
\altaffiltext{5}{
South African Astronomical Observatory, P.O. Box 9, Observatory 7935, South Africa.
}
\altaffiltext{6}{
Department of Astronomy, University of Washington, Box 351580, Seattle, WA 98195.
}
\altaffiltext{7}{
Max Planck Institute for Extraterrestrial Physics, D-85748, Garching, Munich, Germany. 
}
\altaffiltext{8}{
Universitaets-Sternwarte Munich, 1 Scheinerstr, Munich, D-81679, Germany. 
}
\altaffiltext{9}{
Institut de Ci\`encies de l'Espai (IEEC-CSIC), Barcelona, Spain.
}
\altaffiltext{10}{
Department of Physics and Astronomy, Wayne State University, Detroit, MI 48202.          
}
\altaffiltext{11}{
Department of Astronomy and Astrophysics,The University of Chicago, 5640 South Ellis Avenue, Chicago, IL 60637.
}
\altaffiltext{12}{
Center for Astrophysics, Fermi National Accelerator Laboratory, P.~O.~Box 500, Batavia IL 60510. 
}
\altaffiltext{13}{
Institut de F\'{\i}sica d'Altes Energies, Barcelona, Spain.
}
\altaffiltext{14}{
University of Notre Dame, 225 Nieuwland Science, Notre Dame, IN 46556-5670. 
}
\altaffiltext{15}{
The Oskar Klein Centre, Department of Astronomy,  Albanova, Stockholm University, SE-106 91 Stockholm Sweden
}
\altaffiltext{16}{
Department of Physics, Stockholm University, Albanova University Center, S-106 91 Stockholm, Sweden
}
\altaffiltext{17}{
Institute of Astronomy, Graduate School of Science,University of Tokyo 2-21-1, Osawa, Mitaka, Tokyo 181-0015, Japan.
}
\altaffiltext{18}{
Institute of Cosmology and Gravitation,Mercantile House,Hampshire Terrace, University of Portsmouth, Portsmouth PO1 2EG, UK.
}
\altaffiltext{19}{
Instituci\'o Catalana de Recerca i Estudis Avan\c{c}ats, Barcelona,  Spain.
}
\altaffiltext{20}{
Centro de Investigaciones Energ\'eticas, Medioambientales y Tecnol \'ogicas, Madrid, Spain.
}
\altaffiltext{21}{
Space Telescope Science Institute,3700 San Martin Drive, Baltimore, MD 21218.
}
\altaffiltext{22}{
Department of Physics and Astronomy,Johns Hopkins University, 3400 North Charles Street, Baltimore, MD 21218.
}
\altaffiltext{23}{
Department of Physics and Astronomy, University of Pennsylvania, 209 South 33rd Street, Philadelphia, PA 19104.
}
\altaffiltext{24}{
Department of Astronomy and Astrophysics, 525 Davey Laboratory, Pennsylvania State University, University Park, PA 
}
\altaffiltext{25}{
Dark Cosmology Centre, Niels Bohr Institute, University of Copenhagen, Denmark.
}
\altaffiltext{26}{
Department of Astronomy, University of Texas, Austin, TX 78712.
}
\altaffiltext{27}{
Apache Point Observatory, P.O. Box 59, Sunspot, NM 88349.
}

\begin{abstract}
We present measurements of the Type Ia 
supernova (SN) rate in galaxy clusters
based on data from the 
Sloan Digital Sky Survey-II (SDSS-II) Supernova
Survey.
The cluster SN Ia rate is determined 
from 9 SN events in a set of 71 C4 clusters at
$z \le 0.17$ and 27 SN events in 492 maxBCG clusters at $0.1 \le z \le 0.3$.
We find values for the cluster SN Ia rate of
$({\cfsnur}^{+\cfsnurhi+\cfsnurhisyst}_{-\cfsnurlo-\cfsnurlosyst})$ 
$\mathrm{SNu}r$
$h^{2}$ and
$({\bcgsnur}^{+\bcgsnurhi+\bcgsnurhisyst}_{-\bcgsnurlo-\bcgsnurlosyst})$
$\mathrm{SNu}r$
$h^{2}$
($\mathrm{SNu}x = 10^{-12} L_{x\sun}^{-1} \mathrm{yr}^{-1}$)
in C4 and maxBCG clusters, respectively, where the
quoted errors are statistical and systematic, respectively.
The SN rate for early-type galaxies is found to be 
$({\cfsnurearly}^{+\cfsnurhiearly+\cfsnurhisystearly}_{-\cfsnurloearly-\cfsnurlosystearly})$ 
$\mathrm{SNu}r$
$h^{2}$ and
$({\bcgsnurearly}^{+\bcgsnurhiearly+\bcgsnurhisystearly}_{-\bcgsnurloearly-\bcgsnurlosystearly})$
$\mathrm{SNu}r$
$h^{2}$
in C4 and maxBCG clusters, respectively.
The SN rate for the brightest cluster galaxies (BCG) is found to be 
$({\cfsnurbcg}^{+\cfsnurhibcg+\cfsnurhisystbcg}_{-\cfsnurlobcg-\cfsnurlosystbcg})$ 
$\mathrm{SNu}r$
$h^{2}$ and
$({\bcgsnurbcg}^{+\bcgsnurhibcg+\bcgsnurhisystbcg}_{-\bcgsnurlobcg-\bcgsnurlosystbcg})$
$\mathrm{SNu}r$
$h^{2}$
in C4 and maxBCG clusters, respectively.
The ratio of the SN Ia rate in cluster early-type galaxies
to that of the SN Ia rate in field early-type galaxies is
${\cfrat}^{+\cfrathi+\cfratsysthi}_{-\cfratlo-\cfratsystlo}$
and
${\bcgrat}^{+\bcgrathi+\bcgratsysthi}_{-\bcgratlo-\bcgratsystlo}$,
for C4 and maxBCG clusters, respectively.
The SN rate in galaxy clusters as a function of redshift,
which probes the late time SN Ia delay distribution,
shows only weak dependence on redshift.
Combining our current measurements with previous measurements, we 
fit the cluster SN Ia rate data to a linear function of redshift,
and find 
$r_{L} = $
$[(0.49^{+0.15}_{-0.14}) +$ 
$(0.91^{+0.85}_{-0.81}) \times z]$
$\mathrm{SNu}B$ 
$h^{2}$.
A comparison of the radial distribution of SNe in cluster to field early-type
galaxies shows possible evidence for an enhancement of the SN rate in the 
cores of cluster early-type galaxies.
With an observation of at most 3 hostless, intra-cluster SNe Ia, 
we estimate the fraction of cluster SNe that 
are hostless to be  
$(9.4^{+8.3}_{-5.1})\%$.

\end{abstract}

\clearpage
\clearpage

\setcounter{footnote}{0}
\section{Introduction}
\label{sec:clintro}

The rate of Type Ia supernovae (SNe) in 
galaxy clusters is an important field 
of study for a number of reasons.
As discussed in, e.~g.,~\citet{Greggio_05},
supernova (SN) rate measurements are an observational 
probe of the progenitor systems,
with the connection to progenitor models being made through inference of the 
distribution of delay times (DDTs) with respect to star formation.
As galaxy clusters are generally composed of a high fraction 
of early-type galaxies
that have old stellar populations, measurements of the SN Ia rate in clusters 
can in principle simplify the inference of the SN DDT.
SNe in galaxy clusters are also a candidate source for metal
enrichment of the intra-cluster medium (ICM). In particular, 
improved measurement of the rate of intra-cluster SNe 
would be significant for constraining the relative importance of 
sources that may contribute to the cluster ICM enrichment 
(e.~g.~intra-cluster stars vs.~galaxy outflow).

The existing measurements of the cluster SN rate are few
and are generally based on low-number statistics. 
Estimates of the SN cluster rate were first presented by 
\citet{Crane_77} and \citet{Barbon_78}, who considered 
$\approx 5$ SNe discovered in the Coma cluster. 
The cluster SN rate was measured by 
\citet{Galyam_02} in clusters at $z \approx 0.25$ and $z \approx 0.9$.
The \citet{Galyam_02} results are based on a search for SNe in 
archival images of the Hubble Space Telescope ({\it HST}) 
and utilize one and two SNe, 
respectively.
Subsequent to the {\it HST} SN search, a dedicated search for SNe 
in 161 Abell clusters
was undertaken by The Wise Observatory Optical Transient Search (WOOTS) \citep{Galyam_08}.
A total of 6 SNe Ia discovered by the WOOTS were used to determine the 
cluster SN Ia rate at $z\approx 0.15$ by \citet{Sharon_07a}.
A sample of 2-3 SNe Ia from the Supernova Legacy Survey (SNLS) have been 
used to determine the cluster SN Ia rate at $z \approx 0.45$ by \citet{Graham_08}.
Finally, the sample of SNe in the local ($z \lesssim 0.04$) universe 
presented by \citet{Cappellaro_99} have been reanalyzed by 
\citet{Mannucci_08} to determine the cluster SN Ia rate with 
a sample of 12.5 SN Ia (a fractional SN reflects uncertainty in typing; see \citet{Cappellaro_99}).
Additionally, \citet{Mannucci_08} have placed the first constraints on the 
core-collapse (CC) SN rate in galaxy clusters based on a sample of 7.5 CC SNe.
A summary of the SN Ia rate results from the above cluster SN studies is given in Table \ref{tab:clraterev}.
A SN search in 15 massive, high-redshift ($0.5 < z < 0.9$), X-ray selected clusters has been carried out on the 
{\it HST}, as described in \citet{Sharon_07b}, and a measurement of the SN Ia rate 
based on 6-14 SNe discovered by the program is 
forthcoming \citep{Sharon_09}.
A dedicated SN search, targeting $\approx 60 $ X-ray selected clusters in the 
redshift range $0.1 < z <0.2$, 
is also being carried out 
on the Bok 2.~3m telescope on Kitt Peak \citep{Sand_08}

As can be seen in Table \ref{tab:clraterev}, 
the knowledge of the cluster SN Ia rate comprises 
5 measurements, based on a total of $\approx 25$ SNe.
In this paper, we describe new measurements of the cluster SN Ia
rate based on data from 
the Sloan Digital Sky Survey-II (SDSS-II) 
Supernova Survey~\citep{Frieman_08}. The measurements are 
based on 35 SNe 
in the redshift range $0.03 < z < 0.30$, and
therefore represent a significant statistical contribution 
to cluster SN Ia studies. 
In \S \ref{sec:sdsssnobs}
we briefly describe the observations and SN search strategy of the \sns.
In \S \ref{sec:clcats} we describe
the galaxy cluster catalogs employed in this SN rate analysis.
In \S \ref{sec:clsample} we describe selection of the 
cluster SN sample from the \sns~data.
In 
\S \ref{sec:clcorrections}
we describe 
necessary corrections to our SN Ia rate measurements.
In \S \ref{sec:clresults}
we present results on the cluster SN Ia rate, as well as 
limits on the cluster CC SN rate, and studies of the 
distribution of SNe with respect to their host galaxies.
We summarize in \S \ref{sec:clconclude}.
Whenever necessary we assume a flat $\Lambda \mathrm{CDM}$ cosmology with
$\Omega_\mathrm{m} = 1 - \Omega_{\Lambda} = 0.3$, and Hubble 
constant $H_{0} = 70 ~\mathrm{km}~\mathrm{s}^{-1}~\mathrm{Mpc}^{-1}$.

\begin{deluxetable}{rcccrl}
\tablecolumns{6}
\singlespace
\tablewidth{0pc}
\tablecaption{Cluster Rate Measurements
\label{tab:clraterev}
}
\tablehead{
\colhead{Reference} &    
\colhead{Redshift} &    
\colhead{Mean} & 
\colhead{Lookback Time} &       
\colhead{$\mathrm{N}_{\mathrm{SNe}}$} &    
\colhead{SN Ia Rate \tablenotemark{a}} \\

\colhead{} &    
\colhead{Range} &    
\colhead{Redshift} &    
\colhead{[Gyr]} &    
\colhead{} &    
\colhead{[SNuB $h^{2}$]} \\ [-1em]   
}

\startdata
This work (C4)      & 0.03 - 0.17     & 0.084  & 1.11 &   \nsnecf  & $\cfsnub^{+\cfsnubhi+\cfsnubhisyst}_{-\cfsnublo-\cfsnublosyst}$ \\[6pt]
This work (maxBCG)  & 0.10 - 0.30     & 0.225  & 2.69 & \nsnebcgnobias   & $\bcgsnub^{+\bcgsnubhi+\bcgsnubhisyst}_{-\bcgsnublo-\bcgsnubhisyst}$ \\[6pt]
\citet{Mannucci_08} & 0 - 0.04        & 0.020  & 0.28 & 12.5 & $0.57^{+0.22}_{-0.16}$ \\[6pt]
\citet{Sharon_07a}   & 0.06 - 0.19    & 0.150  & 1.89 &   6  & $0.73^{+0.45}_{-0.29}$ \\[6pt]
\citet{Galyam_02}   & $\approx 0.25$  & 0.250  & 2.94 &   1  & $0.80^{+1.84}_{-0.65}$ \\[6pt]
\citet{Graham_08}   & $\approx 0.45$  & 0.450  & 4.67 &   3  & $0.63^{+1.04}_{-0.33}$ \\[6pt]
\citet{Galyam_02}   & $\approx 0.9 $  & 0.900  & 7.30 &   2  & $1.63^{+2.16}_{-1.06}$ \\[6pt]
\enddata
\tablenotetext{a}{
For this work, the quoted errors are statistical and systematic, respectively.
For previous measurements, the total error is quoted; see the 
corresponding references.
}

\end{deluxetable}

\section{SDSS-II Supernova Survey Observations}
\label{sec:sdsssnobs}

Here we briefly describe aspects of the {\sns} most
relevant to the present SN rate analysis. 
Much of the material presented in this section is also relevant to the
SN rate studies described in \citet{Dilday_10},
and is discussed therein; we repeat the discussion here 
for the convenience of the reader. 
The survey is described in more
detail in \citet{Frieman_08} and the SN detection algorithms
are described in \citet{Sako_08}. Additional details
of the survey observations and 
the use of {\it in situ} artificial SNe 
for determining SN detection efficiencies 
are discussed in
\citet{Dilday_08a}.
A technical summary of the SDSS is given by~\citet{York_00}.
Details of the survey calibration 
can be found in~\citet{Hogg_01,Smith_02,Tucker_06}, 
the data processing and quality assessment is described by~\citet{Ivezic_04}, 
and the photometric pipeline is described by~\citet{Lupton_99}

The \sns~was carried out during 
the Fall (September-November) of 2005-2007, 
using the 2.5m telescope \citep{SDSS_telescope} at 
Apache Point Observatory (APO).
Observations were obtained 
in the SDSS $ugriz$ filters \citep{Fukugita_96}
with a wide-field CCD camera \citep{Gunn_98},
operating in time-delay-and-integrate (TDI, or drift scan) mode. 
The region of the sky covered by the 
\sns~(designated stripe 82; see \citet{SDSS_EDR})
was bounded by 
$-60^{\circ} < \alpha_{J2000} < 60^{\circ}$, and 
$-1.258^{\circ} < \delta_{J2000} < 1.258^{\circ}$.
On average any given part of this $\approx 300$ square degree 
area was imaged once every 4 days during the survey operations.

Difference images were produced in the SDSS $gri$ filter bands
by subtracting 
template images, constructed from 
previous observation of the survey region, 
from the survey images, using an implementation of the methods described 
by~\citet{Alard_98}.  
The difference images were searched for positive fluctuations
using the {\tt DoPHOT} photometry 
and object detection package~\citep{Schecter_93};
typical limiting magnitudes ($10 \sigma$ above background) 
for the \sns~were
$g \sim 21.8$, $r \sim 21.5$, and $i \sim 21.2$.
A combination of software cuts and human visual inspection was then used to 
identify promising SN candidates from the full set of transient detections.
As a key component of prioritizing 
SN candidates for spectroscopic observation, 
the light curves for SN candidates were fit to 
models of Type Ia, Type Ib/c and Type II
SNe. 
This procedure is referred to
as {\it photometric-typing}, and is described in detail 
by~\citet{Sako_08}.

Spectroscopic observations,
for both SN type and redshift determination, were provided
by a number of different telescopes. The spectra for the SNe utilized in the 
present SN rate analysis were provided by 
the Hobby-Eberly 9.2m at McDonald Observatory, 
the Astrophysical Research Consortium 3.5m at Apache Point Observatory,
the William-Herschel 4.2m, 
the Hiltner 2.4m at the MDM Observatory, 
the Subaru 8.2m on Mauna Kea,
the 2.6m Nordic Optical Telescope 
and the 3.6m Italian Telescopio  
Nazionale Galileo
at La Palma,
the Mayall 4m telescope at Kitt Peak National Observatory,
and
the 3.5m ESO New Technology Telescope (NTT) at La Silla.
Details of the \sns~spectroscopic data reductions are given by \citet{Zheng_08}.
Comparison to high-quality SDSS galaxy spectra show that SN spectroscopic redshifts 
are accurate to $\approx 0.0005$ when galaxy emission features are used 
and $\approx 0.005$ when SN features are used. In either case the error on the 
spectroscopic SN redshifts are negligible for the SN rate studies considered here.

While the difference imaging pipeline used during the SN search
provides initial photometric measurements, subsequent to the search more 
precise SN photometry is provided 
using a {\it scene modeling photometry} (SMP) technique developed 
by~\citet{Holtzman_08}. 
The final analysis of SN light curves and the selection 
cuts used to define the SN rate sample 
discussed in this paper are based on SMP.

\section{Galaxy Cluster Catalogs and Cluster Luminosity}
\label{sec:clcats}
\begin{figure} [!t] 
\begin{center}
\includegraphics[width=5.75in]{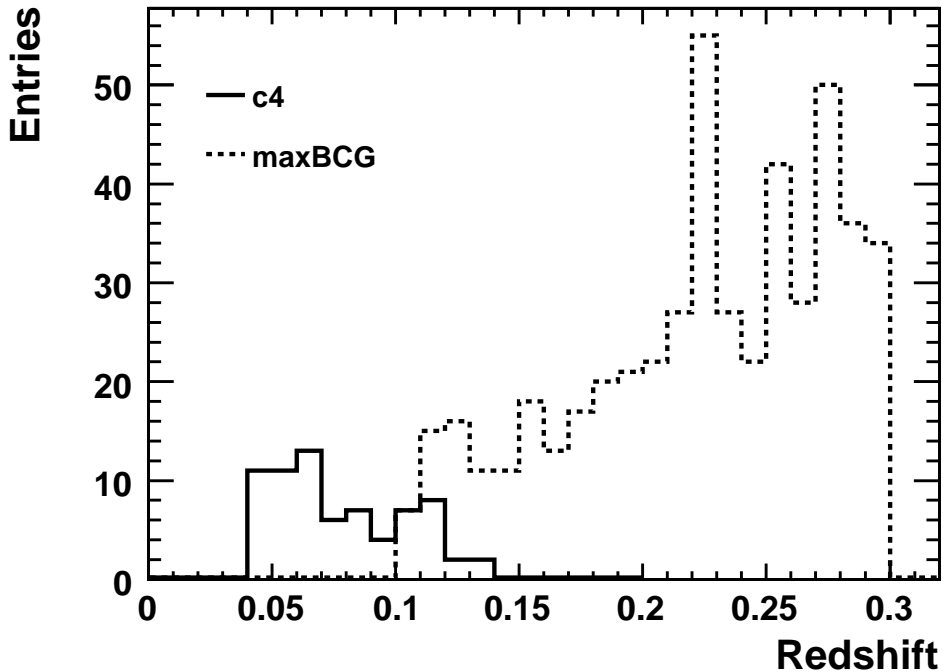}
\end{center}
\caption{
Redshift distributions for the C4 ($ z < 0.17$) and maxBCG ($ 0.1 < z < 0.3$)
clusters in the SDSS-II SN Survey region. 
There are 
71 clusters from the C4 catalog and 
492 clusters from the maxBCG catalog.
 }
\label{fig:fclz}
\end{figure}

In studying the Type Ia SN rate in galaxy clusters we 
will work with two primary cluster catalogs;
the C4 cluster catalog and the maxBCG cluster catalog.
The C4 cluster finding algorithm and catalog are discussed 
in detail by \citet{Miller_05}. 
The maxBCG catalog is presented 
by \citet{Koester_07a}, and the cluster finding 
algorithm is described by \citet{Koester_07b}.
We briefly describe and summarize the 
content of these two cluster catalogs below.
The redshift distributions 
for the clusters
in these two catalogs that are in the SDSS-II Supernova Survey region
are shown 
in Figure~\ref{fig:fclz}.

\subsection{The C4 Cluster Catalog Description}
\label{sec:c4descript}

The C4 cluster catalog is based on the main spectroscopic sample of the 
SDSS and contains clusters in the redshift range $ 0.03 < z < 0.17$. 
The C4 cluster catalog provides, among other quantities,
cluster coordinates, 
the luminosity of each cluster in the SDSS $r$-band, and
the number of galaxies identified as members of each cluster, $N_{\mathrm{gals}}$.
The main spectroscopic sample of galaxies from SDSS is designed to be 
complete to a limiting magnitude of $r \approx 17.8$ \citep{Strauss_02}. 
For typical cluster galaxy luminosities, this implies that the 
identification of member galaxies, with $L > 0.4~L_{\star}$, 
is complete for clusters at $z < 0.11$. 
For clusters above this redshift limit a correction has to be made to the 
total cluster luminosity.
The C4 cluster identification algorithm works by 
searching for groups of objects that are tightly 
clustered in a 7-dimensional feature space, that includes 
spatial position, redshift, and observed colors. Note that there
is no requirement that the colors for the galaxies be consistent
with the colors of early-type galaxies, only that they be 
consistent with one another.
The SDSS main galaxy sample cannot be $100\%$ complete, 
as the SDSS fiber-spectrograph imposes a minimum
angular separation of 55'' ($\approx 100$ kpc $h^{-1}$ at $z=0.15$) 
for objects targeted for the SDSS main 
spectroscopic galaxy sample \citep{SDSS_EDR}. A correction is applied to 
the C4 cluster luminosities to account for this by including 
galaxies as members of the cluster when they satisfy the 
magnitude requirement ($r < 17.8$) and have similar colors
to the spectroscopically determined cluster members.
The published C4 catalog is based on the 2nd data release of the 
SDSS \citep{SDSS_DR2} and contains 748 clusters. The catalog used
in this work is an extended version of the C4 catalog that contains 
1713 clusters and is 
based on the 5th data release of the SDSS \citep{SDSS_DR5}
(B. Nichol, private communication).
The subset of clusters that we will use in measuring the cluster 
SN rate are those that lie within stripe 82, and that have 
$ -40^{\circ} < ~\alpha_{\mathrm{J2000}} < 50^{\circ}$.
There are 71 C4 clusters in this subset.

\subsection{The maxBCG Cluster Catalog Description}

\begin{figure} [!t] 
\begin{center}
\includegraphics[width=5.75in]{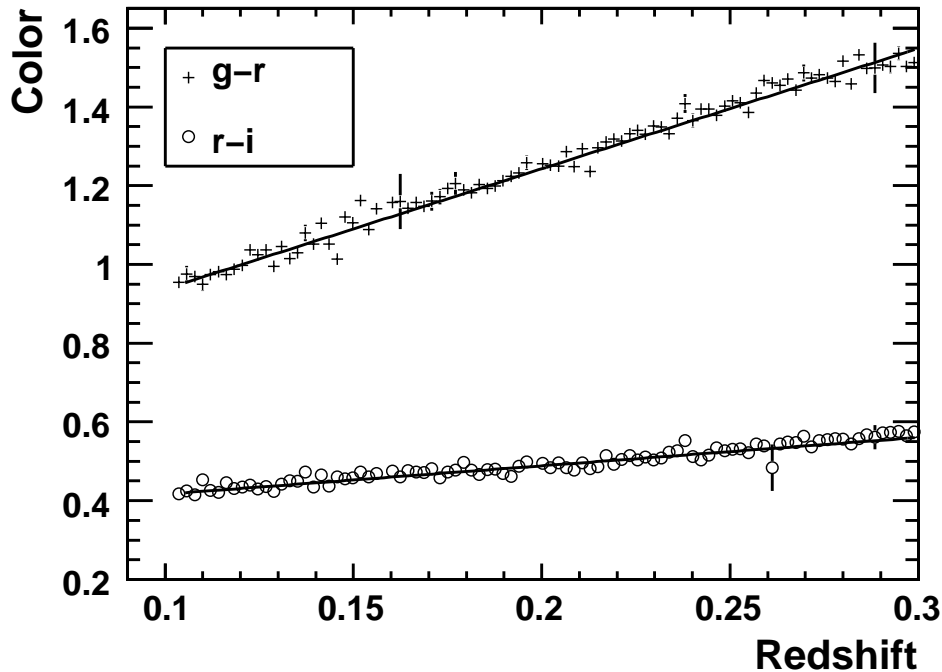}
\end{center}
\caption{
Observed colors of maxBCG galaxies vs.~redshift.
The points represent the mean at the corresponding
redshift, and the error bars represent the 
error on the mean. 
The solid lines show the best fitting linear function.
 }
\label{fig:bcgcolorsz}
\end{figure}

The maxBCG catalog is based on SDSS photometric measurements, and the 
cluster identification algorithm relies on the tight relationship
between color and redshift for luminous red galaxies, which make
up the majority of galaxy cluster composition. The maxBCG 
algorithm assigns a photometric redshift to each identified
cluster that is derived by comparing the cluster member galaxy
colors to the expected colors for early-type galaxies, as a function
of redshift. Comparison of the photometric redshifts for the 
maxBCG clusters to the spectroscopically measured redshift of the 
brightest cluster galaxy (BCG), when available, shows that 
the residuals for redshift (photometric - spectroscopic) 
are well described by a Gaussian distribution with a mean 
of 0 and a 
standard deviation
of $\sigma \approx 0.015$.
The observed ($g-r$) and ($r-i$) colors for maxBCG member galaxies, as 
a function of redshift, are shown 
in Figure \ref{fig:bcgcolorsz}.
Linear functions were fit to to the $(r-i)$ and $(g-r)$ 
colors as a function of redshift, $z$, and these will be used below 
for applying k-corrections to the observations.
For reference, the functions derived, valid for $0.1 \le z \le 0.3$, are,
\begin{eqnarray}
(r-i) & = & 0.345 + 0.720~z  \\
(g-r) & = & 0.632 + 3.054~z.  
\end{eqnarray}

The maxBCG catalog
is restricted to the redshift interval $ 0.1 < z < 0.3$,
and is thus highly complementary to the C4 catalog. The lower-limit 
for the maxBCG cluster catalog is imposed because the 
fractional photometric redshift errors for redshifts 
less than $z=0.1$ are significant and have a
large systematic effect on derived cluster properties.
Furthermore, cluster catalogs based on spectroscopic 
redshifts for $z<0.1$ are already available. 
The upper limit is imposed because, at $z \approx 0.3$, the
``$4000~\AA$ break'' that is responsible for the uniformity in 
early-type galaxy colors moves into the region between the 
SDSS observer frame $g$ and $r$ band filters. Thus, the 
accuracy and precision for galaxy photometric redshifts
is severely diminished \citep{Koester_07a}. 
A lower-limit for the luminosity of cluster members is imposed 
so that the definition of the composition of the clusters
is consistent across the redshift range.
The limit corresponds to an absolute magnitude of
$ ^{(0.25)} i \approx -20.25$, 
(the $^{(0.25)} i$ notation is explained in \S \ref{sec:cllumcontent})
and is such that the catalog is volume
limited over the entire redshift range $ 0.1 < z < 0.3$.
Measures of the physical extent, $r_{200}$, and of the richness, $N_{200}$ and $ L_{200}$, are provided for each cluster.
$r_{200}$ is defined as the radius such that the 
mean density of early-type galaxies contained within is 200 times
greater than the mean density of such galaxies \citep{Koester_07a}; $N_{200} (L_{200})$ is defined as the 
the number (total luminosity) of early-type galaxies contained within $r_{200}$.
The public maxBCG cluster catalog contains 13,823 clusters with 
$N_{200} \ge 10$.
The subset of maxBCG clusters that we will use in measuring the cluster 
SN rate are those that lie within stripe 82, and that have 
$ -50^{\circ} < ~\alpha_{\mathrm{J2000}} < 60^{\circ}$.
There are 492 maxBCG clusters in this subset. Additionally, this
work makes use of a catalog of the maxBCG member galaxies 
(B. Koester, private communication).

In addition to the cluster richness and luminosity 
estimates that are provided with the maxBCG catalog, 
the maxBCG clusters have been extensively 
studied by e.g.~\citet{Sheldon_09a, Johnston_07, Sheldon_09b}.
Of particular relevance to the study of the SN rate in galaxy clusters 
are the luminosity functions (LFs) of maxBCG clusters presented
by \citet{Hansen_07}, and we will make extensive use of these below.

\subsection{Luminosity Content of C4 and maxBCG Cluster Catalogs}
\label{sec:cllumcontent}

In \S \ref{sec:clresults} we will present the SN rate 
in galaxy clusters per unit luminosity. 
In this section we present a comprehensive discussion
of the luminosity content for 
the C4 and maxBCG cluster catalogs. 
In what follows we use the notation $^{(z)} m$
to denote the observer frame magnitude, $m$, for an object
that has been k-corrected to a redshift $z$.
All luminosities presented here are for galaxies with 
$L > 0.4~L_{*}$ ($L_{*}$ is a 
characteristic luminosity for cluster members; see
Equation \ref{eqn:lfschect}). Correcting for the faint end of the 
luminosity distribution is discussed below.

The C4 catalog provides total cluster luminosities in 
SDSS $r$-band. The total {\it uncorrected} $r$-band luminosity
for the galaxies identified as  
cluster members, for clusters considered in this study, is 
$2.02 \times 10^{13} ~L_{\sun} ~h^{-2}$, where 
$L_{\sun}$ is the luminosity of the sun, and $h$ is 
the value of the Hubble constant 
in units of $100 ~\mathrm{km} ~\mathrm{s}^{-1} ~\mathrm{Mpc}^{-1}$.
As discussed in \S \ref{sec:c4descript}, these luminosities
have to be corrected to account for incompleteness
of the spectroscopic sample due to fiber-collisions. 
The total $r$-band luminosity {\it after} making this
correction is also provided by the C4 catalog and has the value
$4.08 \times 10^{13} ~L_{\sun} ~h^{-2}$. 
A mentioned above, the luminosities for clusters at 
$z > 0.11$ also
have to be corrected to account for the fact that 
some cluster galaxies with $L > 0.4 L_{\star}$ will have observed magnitudes
fainter than the completeness limit of the main SDSS galaxy 
sample ($r \approx 17.8$). Of the 71 C4 clusters 
we are considering, 12 are at $z > 0.11$.
Applying a correction to the luminosities of these 12 clusters,
assuming the cluster luminosity functions of \citet{Hansen_07},
results in a total $r$-band luminosity for C4 clusters in this study
of $4.12 \times 10^{13} ~L_{\sun} ~h^{-2}$. 

The maxBCG catalog includes the summed luminosities 
for member galaxies in $^{(0.25)} i$ and $^{(0.25)} r$.
However, the maxBCG cluster LFs presented by 
\citet{Hansen_07} represent a more complete study
of the luminosity content of maxBCG clusters, and
we will use these as the definitive measure of 
the total maxBCG cluster luminosities. In particular, 
a background subtraction has been performed that 
reduces inaccuracies in the cluster luminosities
due to interloping foreground and background 
galaxies that may be counted as maxBCG cluster members.
In \citet{Hansen_07}, luminosity functions (LFs) are 
presented for maxBCG satellites, as 
a function of the richness measure, $N_{200}$.
The LFs account for both red and blue cluster galaxies, but 
do not include the contribution to the luminosity from the BCG.
The LFs are assumed to take the form of a Schechter function,

\begin{equation}
\label{eqn:lfschect}
\phi(L) ~dL = \phi^{*} ~\left(\frac{L}{L_{*}}\right)^{\alpha} ~e^{-L/L_{*}} ~\frac{dL}{L_{*}}
\end{equation}

\noindent where $L_{*}$ is a characteristic luminosity for 
cluster members, and $\phi^{*}$ is a normalization constant
with units of inverse volume. The 3 parameters of the 
LFs, $\phi^{*}, L_{*}, \alpha$, are each expressed 
as functions of $N_{200}$, with the general functional form of 
$A \times N_{200}^{\beta}$.
The values of A (normalization) 
and $\beta$ (exponent) for these 3 LF parameters are
given in Table \ref{tab:clLF}.
To use the LFs to compute the total luminosity we exploit the identity
\begin{equation}
\label{eqn:sigl}
\Sigma ~L = \langle L(N_{200}) \rangle ~N_{200}
\end{equation}

\noindent where $\Sigma L$ denotes the summed luminosity for the cluster, 
and the average luminosity, $\langle L \rangle$, is given by,

\begin{equation}
\label{eqn:avgl}
\langle L(N_{200}) \rangle = 
\frac{\int_{0.4 L_{*}}^{\infty} dL ~L ~\phi(L)}
{\int_{0.4 L_{*}}^{\infty} dL ~\phi(L)}.
\end{equation}

\noindent Equation \ref{eqn:sigl} gives the total $^{(0.25)} i$
luminosity in a cluster, as a function of $N_{200}$. The total 
luminosity in clusters in the SN survey region is then
the sum over $N_{200}$ of $\langle L(N_{200}) \rangle ~N_{SN}(N_{200})$,
where $N_{SN}(N_{200})$ is the number of 
clusters in the survey region 
for which the number of member galaxies is $N_{200}$.
Using this formalism we find that the total luminosity of maxBCG clusters
in the survey region 
is $\Sigma L^{0.25}_{i} = 1.096 \times 10^{14} ~L_{\sun} ~h^{-2}$.

\begin{deluxetable}{ccr}
\tablecolumns{3}
\singlespace
\tablewidth{0pc}
\tablecaption{maxBCG Luminosity Function Parameters
\label{tab:clLF}
}
\tablehead{
\colhead{LF Parameter} &    
\colhead{Normalization} &
\colhead{Exponent} \\
\colhead{} &
\colhead{(A)} &
\colhead{($\beta$)} 
}

\startdata
$\phi^{*}$ &  $8.0 ~\mathrm{Mpc}^{-3} ~h^{3}$   & $-$0.20 \\
$L_{*}$    &  $0.8 \times 10^{10} ~L_{\sun} ~h^{-2} $ &  0.15 \\
$\alpha$   & $-$0.28  &  0.25 \\
\enddata

\tablecomments{Parameters refer to Schechter functions 
derived by \citet{Hansen_07}.}
\end{deluxetable}

\subsubsection{Correcting Cluster Luminosities for Faint Galaxies}
The luminosities quoted above for the C4 and maxBCG cluster 
catalogs include only galaxies with $L > 0.4~L_{*}$, which is
a conventional way of characterizing cluster luminosities. 
In measuring the cluster SN rate we do not wish to 
exclude SNe occurring in faint galaxies, and so it is 
necessary to estimate the contribution to the
total cluster luminosities from galaxies with $L < 0.4~L_{*}$.
The total cluster luminosity can be estimated as
$L = \kappa ~L^{+}$, where $L^{+}$ denotes the luminosity 
for galaxies with $L > 0.4~L_{*}$ and the correction factor, $\kappa$, is 
given by
\begin{equation}
\kappa = \frac
{\int_{0}^{\infty} dL ~L ~\phi(L)}
{\int_{0.4 L_{*}}^{\infty} dL ~L ~\phi(L)}.
\end{equation}

\noindent For a typical maxBCG cluster with $N_{200} = 20$ the 
power-law exponent of the luminosity function is 
$\alpha = -0.59$ (Table \ref{tab:clLF}). For
a cluster with $N_{200} = 35$ the  
value for the exponent is $\alpha = -0.68$.
The corresponding correction factors 
are $\kappa = 1.21$ 
and 
$\kappa = 1.25$. 
We will assume that the faint end of the luminosity function
is a characteristic property of galaxy clusters and that 
the power-law behavior of the LFs for maxBCG clusters
is appropriate for the C4 clusters also.

\subsubsection{k-corrections}
\label{sec:kcors}
\begin{figure} [!t] 
\begin{center}
\includegraphics[width=5.75in]{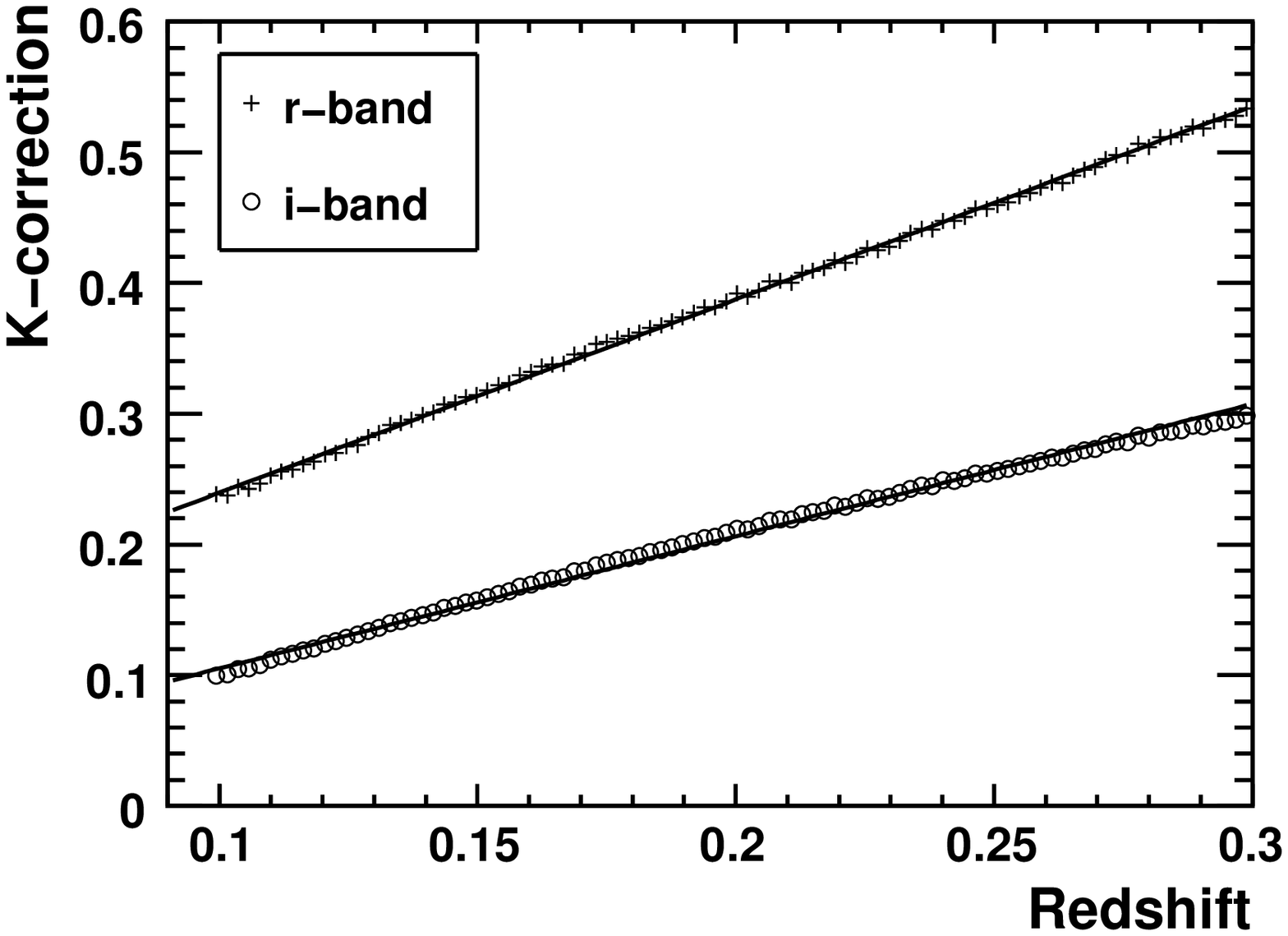}
\end{center}
\caption{
K-corrections vs.~redshift for early-type galaxies.
The points represent the mean at the corresponding
redshift, and the error bars represent the 
error on the mean. 
The solid lines show the best fitting linear function.
 }
\label{fig:bcgkcorz}
\end{figure}

In order to minimize uncertainties due to k-corrections, 
the maxBCG magnitudes, and the corresponding luminosities, 
have been 
k-corrected to the median redshift of the clusters, $ z = 0.25$.
For comparison of the SN rate results based on the 
maxBCG clusters with previous cluster SN rate 
measurements, it will be necessary to k-correct the luminosities
into more standard filters. 
In order to determine the appropriate k-corrections we
selected a set of galaxies from the SDSS galaxy catalog
that satisfy the color vs.~redshift relations 
for maxBCG members discussed above. The corresponding 
set of galaxies contains $\approx 675,000$ members.
We then matched these galaxies with their 
corresponding records in the {\tt photoz} database provided by the SDSS 
catalog archive server (CAS),
which provides k-corrections 
based on the work of \citet{Blanton_03a}. We thereby determine 
an average k-correction as a function of redshift, appropriate to the
early-type galaxies that make up the bulk of the maxBCG catalog. 
The k-corrections so derived, in SDSS $r$ and $i$ bands, 
are shown in Figure \ref{fig:bcgkcorz}.
For reference, the functions employed for k-corrections as a function
of redshift, $z$, valid for $0.1 \le z \le 0.3$, are
\begin{eqnarray}
K_{rr}(z) & = & 9.17 \times 10^{-2} + 1.48~z  \\
K_{ii}(z) & = & 3.95 \times 10^{-3} + 1.01~z.
\end{eqnarray}

\noindent The functions above were derived by binning the k-corrections
in redshift and taking the error on the mean 
(root mean square divided by square root of the number of entries) as the uncertainty.
With this formalism the error on the fitted parameters is $\lesssim 1\%$ and is thus
negligible in comparison to the statistical error on our SN Ia rate measurements.

To transform the maxBCG cluster luminosities from $^{(0.25)} i$
to $r$, we note the following identity:
\begin{eqnarray}
m_{r} & = & m^{0.25}_{r} - K_{rr}(z=0.25) \\
      & = & m^{0.25}_{i} + (r-i)(z=0.25) - K_{rr}(z=0.25)
\end{eqnarray}

\noindent We k-correct the maxBCG cluster luminosities 
from $^{(0.25)} i$ to $r$ because at a redshift $z=0.25$ 
rest-frame $r$ maps closely to observer-frame $i$, and 
because the analysis of \citet{Hansen_07} that we
take as the definitive measurement of the maxBCG cluster 
luminosities is performed 
for $^{(0.25)} i$ and not $^{(0.25)} r$. 
Using the expressions given above 
for $(r-i)$ and $K_{rr}$, evaluated at $z = 0.25$, we
derive the transformation $m_{r} = {^{(0.25)}m_{i}} + 0.063$.
For comparison to other cluster SN rate measurements, 
it is necessary to express the cluster luminosities
in units of $L^{r}_{\sun}$.
The conversion to solar luminosities is given by

\begin{equation} 
\label{eqn:ltransf}
\left(\frac{L}{L_{\sun}}\right)_{r} 
= \left(\frac{L}{L_{\sun}}\right)_{^{0.25}i} 
\times 10^{-0.4 
( (m_{r} - m_{^{0.25}i}) -  
(M^{r}_{\sun} - M^{^{0.25}i}_{\sun}))},
\end{equation}

\noindent where $M_{\sun}$ is the absolute magnitude of the sun.
To compute the absolute magnitude of the sun in an arbitrary 
filter, we use a solar spectrum obtained from the 
CALSPEC\footnote{http://www.stsci.edu/hst/observatory/cdbs/calspec.html} database, hosted by the Space Telescope Science Institute,
and compute synthetic magnitudes using a custom piece
of software written for this task.
We thereby derive values for the absolute magnitude of the sun
of $M^{r}_{\sun} = 4.64$ (in agreement with the value
published by \citet{Blanton_07}), and 
$M^{^{0.25}i}_{\sun} = 4.67$.
With these values, Equation \ref{eqn:ltransf} becomes
$(L/L_{\sun})_{r} = 0.918 ~(L/L_{\sun})_{^{0.25}i}$, 
giving the $r$-band luminosity in maxBCG clusters, for galaxies
with $L > 0.4~L_{*}$, as 
$\Sigma L_{r} = 1.006 \times ~10^{14} ~L_{\sun}$.

\subsection{Uncertainty on the Cluster Luminosities}
\label{sec:clust_uncert}
Any uncertainty on the luminosity of the clusters described above 
will propagate to an uncertainty on the derived cluster SN Ia rates. 
Sources of uncertainty on the cluster luminosities might include, e.g.,
uncertainties in the background subtraction used in defining the 
maxBCG cluster luminosities, or uncertainty in the correction 
for fiber collisions in defining the C4 clusters.
The cluster catalogs described above do not include uncertainties
on the cluster luminosities, and it is beyond the scope of this
analysis to determine these uncertainties. 
However, as an estimate the order of magnitude of the 
cluster luminosity uncertainties, we note that 
\citet{Menanteau_10} give uncertainties on the luminosities of 
individual clusters, for a cluster catalog that was constructed 
in a similar way to the maxBCG cluster catalog. 
These uncertainties are $\approx 10\%$, and include
both statistical and systematic effects. As discussed above, 
the SN Ia rate analysis includes 71 and 492 clusters from 
the C4 and maxBCG catalogs, respectively. If the uncertainties on the 
individual clusters are $\approx 10 \%$ and are independent,
then the uncertainty on the total cluster luminosities will
be $\lesssim 1\%$. We conclude that uncertainty on the cluster 
luminosities is likely to be negligible in comparison to the 
statistical uncertainty on the cluster SN Ia rate measurement, but
we acknowledge that any additional uncertainties on the 
cluster luminosities, or correlations between the uncertainties on the
luminosity of the individual clusters, would impact the precision of our 
cluster SN Ia rate measurements.

\section{The Cluster SN Sample}
\label{sec:clsample}

\subsection{Type Ia Supernovae}
\label{sec:clsampleia}
To define the SN sample that is associated with galaxy 
clusters it is necessary to define selection criteria on both
the SN light-curve properties, and the SN spatial correlation 
with clusters. 
As was done in our study of the low redshift SN rate from the first season
of the \sns~\citep{Dilday_08a}, the SNe in the rate
sample are required to meet certain selection criteria on their
light-curve fits to the MLCS2k2 light curve model \citep{Jha_07}. 
For the present analysis the selection criteria are,

\begin{itemize}
\item At least 1 observation at $< -2$ days relative to peak
in the supernova rest-frame.

\item At least 1 observation at $> \mathrm{t}_{\mathrm{late}}$ days
relative to peak in the supernova rest-frame.

For maxBCG cluster SN candidates
$\mathrm{t}_{\mathrm{late}} = 10$ days, and for C4 cluster SN candidates
$\mathrm{t}_{\mathrm{late}} = 5$ days.  
The peak of the SN light curve is defined as the 
epoch of maximum luminosity in the rest-frame $B$-band.

\item Fit probability $> 10^{-3}$. 

The fit probability is computed assuming that the value of the minimum 
$\chi^2$ from the MLCS2k2 light curve fit follows a $\chi^2$ distribution
with $N_p - 4$ degrees of freedom, with $N_p$ the number of photometric
observations. All SNe light curves are fit for epoch of maximum light,
luminosity parameter ($\Delta$) and extinction parameter ($A_V$).
When the SN redshift is known from spectroscopic measurement, the 
SN light curve is additionally fit for distance modulus. In the 
case of photometric SNe, the additional fit parameter
is the redshift of the SN.

\end{itemize}

\noindent The requirement of a late time photometric measurement
is used primarily to reduce misidentification of photometric SN Ia
candidates. 
For the low redshifts of the C4 galaxy clusters the majority of SNe 
have spectroscopic confirmation of type, and so a requirement of 
$t_{\mathrm{late}} > 5$ days increases the size of the SN sample, 
without introducing uncertainty from misidentification of 
photometric SNe. For the maxBCG clusters, the number of  
photometric cluster SN candidates is significant and a stricter 
requirement, $t_{\mathrm{late}} > 10$ must be used.
Additionally, the SNe that will be used for determining the SN rate 
are required to satisfy a set of selection criteria, based on fits of 
their search photometry to models of Type Ia, Type II and Type Ib/c light curve models.
These additional selection criteria are identical to those discussed in more detail in \citet{Dilday_10}. Briefly, 
the SN candidates must be discovered in at least 3 epochs of the SN search, must fit the Type Ia model well
in relation to the Type II and Type Ib/c models, and must not have an overly broad light curve shape, as typified by the
peculiar SN Ia 2005gj \citep{Aldering_06,Prieto_07}. 
The combination of cuts on light-curve sampling and shape 
effectively reject AGN from the cluster SN Ia sample.

A SN candidate is defined to be associated with a galaxy cluster 
if it satisfies
\begin{itemize}
\item SN is within 1 Mpc $h^{-1}$ projected distance of the center of a cluster.
\item SN redshift is consistent with the cluster redshift. 
\end{itemize}

\noindent Projected distance here refers to the orthogonal distance 
from the SN to the center of the galaxy cluster, assuming the
redshift of the cluster.
The choice of 1 Mpc $h^{-1}$ projected distance to
define association with a cluster is chosen 
largely for consistency with previous cluster 
SN Ia rate measurements.
The definition of redshift consistency depends on whether the
SN and/or cluster redshifts are determined photometrically
or spectroscopically.
For the C4 clusters, the cluster redshifts are always 
precisely determined with a spectroscopic measurement,
whereas the maxBCG cluster redshifts are determined 
photometrically. Consistency between the 
SN redshift, $z_{s}$, and cluster redshift, $z_{c}$,  
is defined in the following way:

\begin{itemize}
\item Spectroscopic SN redshift and spectroscopic cluster redshift \\
$| z_{s} - z_{c} | < 0.015$
\item Spectroscopic SN redshift and photometric cluster redshift \\
$| z_{s} - z_{c} | < 0.025$
\item Photometric SN redshift and spectroscopic cluster redshift \\
$| z_{s} - z_{c} | < 2.5 ~\sqrt{(0.01)^2 + \delta z_{s}^2}$
\item Photometric SN redshift and photometric cluster redshift \\
$| z_{s} - z_{c} | < 2.5 ~\sqrt{(0.015)^2 + \delta z_{s}^2}$
\end{itemize}

\noindent where $\delta z_{s}$ is the error on the SN 
photometric redshift.
In the case that both the SN and the cluster have spectroscopically
determined redshifts, the allowable spread in galaxy redshifts 
is taken as 0.015. This number was determined from the 
catalog of maxBCG cluster member galaxies. For each cluster that has
at least 4 member galaxies with spectroscopically 
determined redshifts, we computed the root mean square (RMS) of
the difference between the member spectroscopic redshifts and the redshift
assigned to the cluster. The mean of the resulting distribution
of RMS values is $\approx 0.015$. A similar value is 
obtained if the difference between the 
cluster redshift and the member galaxy spectroscopic redshifts
are instead fit with Gaussian distributions, and the mean of the distribution
of RMS values for each Gaussian is computed. 
We note that for the 9 C4 cluster SNe that have spectroscopic 
redshifts for both the SN and the cluster, the deviation in 
redshift is $\lesssim 0.002$ and this cut could be made more
strict without affecting the SN rate result. In the case 
that the SN has a spectroscopically measured redshift and the 
cluster has a photometrically determined redshift (maxBCG clusters only)
the tolerance on redshift consistency (0.025) is determined by 
adding in quadrature the RMS spread of maxBCG member galaxy redshifts 
mentioned above (0.015)
and the typical accuracy of the maxBCG 
cluster photometric redshifts ($\approx 0.015- 0.020$).

In the case that the SN redshift is determined photometrically,
the cuts we apply correspond to a tolerance of $2.5~\sigma_z$, 
where $\sigma_z$ represents the error on the SN redshift, added in 
quadrature to a tolerance on the cluster redshift. This is
a rather loose tolerance on redshift consistency,
and we discuss this issue further in \S \ref{sec:clcorrections}.
There are 9 SNe Ia in C4 clusters and 27 SNe Ia from maxBCG 
clusters that satisfy the selection criteria, and these 
are listed in 
Table \ref{tab:snec4} (C4 clusters) and 
Table \ref{tab:snebcg} (maxBCG clusters), respectively. 
We note that SNe 12979 and 18375 occur in a cluster that 
is a member of both the C4 and maxBCG catalogs, and 
thus there are 34 distinct SNe Ia that pass the selection criteria.
We note that SNe 16280, 18047 and 18362 are each associated with 
two distinct clusters, and
that SNe 14279 and 16215 are associated with the same cluster. In the
cases where a SN is associated with two clusters, we will count it 
only once in the SN rate calculation, assigning it to the 
nearest of the two clusters.

\section{Corrections to the SN Ia Rate Measurements}
\label{sec:clcorrections}

\subsection{Search Efficiency}
\label{sec:cleffs}

\begin{figure} [!t] 
\begin{center}
\includegraphics[width=5.75in]{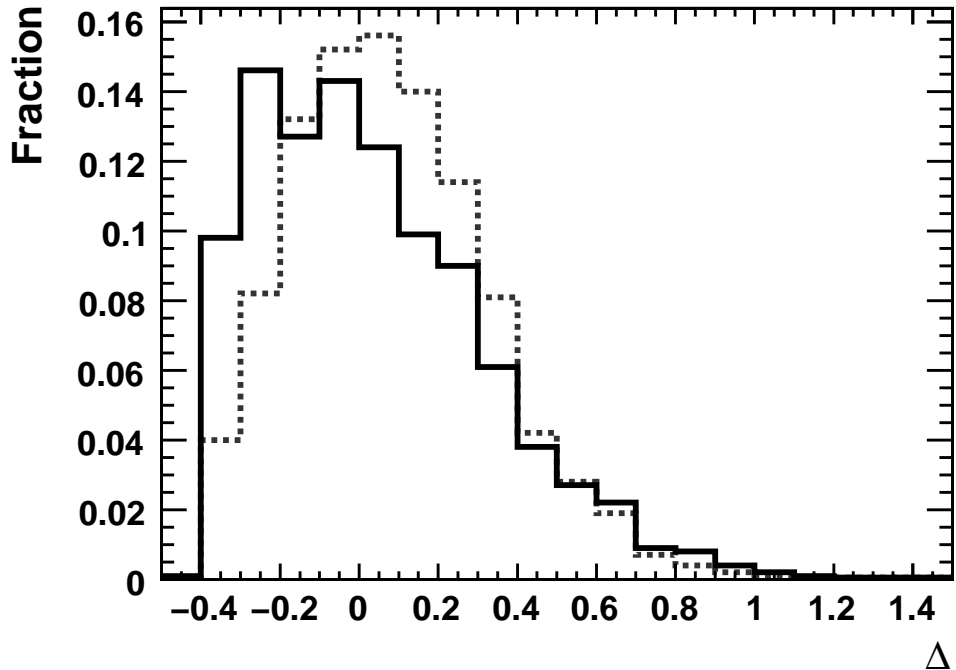}
\end{center}
\caption{
Assumed distribution for the MLCS2k2 $\Delta$ parameter for the
entire SN sample (black/solid) and for the intrinsically fainter
cluster SNe (gray/dashed).
}
\label{fig:cldeltadist}
\end{figure}

The method for determining the SN discovery efficiency 
is based on the same Monte Carlo (MC) studies discussed in \citet{Dilday_10}.
However there is an important modification to 
be considered for SNe in galaxy clusters.
It is a well established result that Type Ia
SNe in early-type galaxies are more
likely to be intrinsically faint, fast-declining 
SNe (e.~g.~\citet{Sullivan_06, Jha_07, Smith_09}). Therefore, the
assumed distribution for the luminosity parameter of the 
MLCS2k2 light curve model, $\Delta$,  for the 
entire SN sample, as employed in 
\citet{Dilday_10}, is not appropriate for considering 
SNe in galaxy clusters. 
To determine the 
SN discovery efficiency for SNe in galaxy clusters, 
we generated a
set of MC SNe with a $\Delta$ distribution
similar to the observed distribution for cluster SNe.
The $\Delta$ distribution assumed for 
cluster SNe Ia is shown in Figure \ref{fig:cldeltadist}.
The fiducial $\Delta$ distribution has mean 
$\langle \Delta \rangle = 0.033$, 
with an RMS of $0.286$, 
and the cluster SNe $\Delta$ distribution has mean
$\langle \Delta \rangle = 0.081$, 
with an RMS of $0.246$.
Assuming a larger mean value for $\Delta$ 
does not significantly affect the 
SN selection efficiency 
at the low redshifts of the C4 clusters, 
and has a $2\%$ effect at the
redshifts of the maxBCG clusters.
The SN selection efficiency at the redshifts of the C4 clusters
is approximately constant for each observing season, with the 
values 0.77, 0.73 and 0.72 for the 2005, 2006 and 2007
observing seasons, respectively.
The redshift dependent efficiencies used for the 
maxBCG cluster SNe are discussed in \citet{Dilday_10}.
Briefly, the efficiencies are well described by a function,
$\epsilon_{0}/(1+e^{(z - z_0)/s_z})$, 
with 
$\epsilon_{0} \approx 0.7$,
$z_{0} \approx 0.35$ and 
$s_{z} \approx 0.05$.
As in \citet{Dilday_08a} and \citet{Dilday_10} the 
efficiency at low-redshift is significantly less than 1 due 
mainly to the requirements of early and late time observing
epochs.

\subsection{Contamination From Redshift Uncertainties}
\label{sec:clphotbias}

In \S \ref{sec:clsample}, we have defined a set of criteria on the consistency 
of the SN and cluster redshifts, which, in the case of photometric redshifts, 
correspond roughly to a requirement that the redshifts match to within
$\approx 2.5 \sigma$. We chose to apply a loose cut on the photometric 
redshift consistency because cluster SNe are rare by their nature
and we wish to maximize the size of the sample, and because we assert that
the requirement of a spatial correspondence with a cluster is 
already a strong filter on contamination from chance projections.
Here we quantify further the expected contamination rate from 
chance projections, in the presence of finite error on the SN redshift 
measurement. We note that for the C4 cluster SN sample,
all the SNe have spectroscopically measured redshifts 
(see Table \ref{tab:snec4}), such that the 
contamination rate due to using SN photometric redshifts is 
manifestly 0, and we focus exclusively on contamination in the maxBCG SN sample.

To estimate the expected contamination rate we employ the following 
procedure. For each cluster in the maxBCG catalog we compute the 
expected contamination, 

\begin{equation}
n^{i}_{\mathrm{contam}} =
\int_{0}^{\infty} dz~\eta(z)
\int_{z_{\mathrm{cl}}-n_{\sigma}\sigma_z}^{z_{\mathrm{cl}}+n_{\sigma}\sigma_z}~dz' 
\frac{1}{\sqrt{2\pi}\sigma_z} 
\exp(\frac{-(z-z')^2}{2\sigma_z^2}),
\end{equation}

\noindent where 
$\eta(z)$ is the SN number density, per unit redshift,
$z_{\mathrm{cl}}$ is the cluster redshift, 
$\sigma_z$ is the assumed error on the SN photometric redshift,
and $n_{\sigma}$ defines our cut on redshift consistency.
The error on the photometric redshift, as a function of 
redshift, is shown in \citet{Dilday_10} to be 
$\sigma_z \approx 0.2 \times z^{1.5}$. 
To compute $\eta(z)$, we use the power-law
rate model from \citet{Dilday_08a} 
($r_V = 2.6 \times 10^{-5}~(1+z)^{1.5}$
$\mathrm{SNe}$
$\mathrm{Mpc}^{-3}$
$h_{70}^{3}$), along with the 
\sns~SN efficiency function. The solid angle
used in computing $\eta(z)$ is the solid angle
corresponding to a 1 Mpc $h^{-1}$ separation
from a cluster at redshift $z_{\mathrm{cl}}$.
However, we assume that no SNe will be 
spatially associated with a cluster and have a 
redshift within $\pm 0.015$ of the cluster
unless it is, in fact, associated with the cluster. Therefore
$\eta(z)$ is assumed to be 0 for 
$-0.015 < 
z-z_{\mathrm{cl}}
< +0.015$. 
The choice of $0.015$ as the spread of maxBCG cluster redshifts is
discussed in \S \ref{sec:clsampleia}.
For each cluster in the maxBCG catalog, we also
compute the number of expected cluster SNe Ia, $N^{T}$,
assuming a cluster SN Ia rate per unit $r$-band luminosity that
is broadly consistent with those given by 
\citet{Sharon_07a, Mannucci_08}.
The observed number of SNe that would be included in the sample,
$N^{O}$,
is then related to the true number, $N^{T}$, by

\begin{equation}
\label{eqn:clcontam}
\frac{N^{O}}{N^{T}} = 
\mathrm{erf}(n_{\sigma}/\sqrt(2)) + 
\frac{N_{\mathrm{contam}}}{N^{T}}
\end{equation}

\noindent where erf(.) is the error function, and
$N_{\mathrm{contam}}$ is the total expected contamination, 
$\Sigma_{i}~n^{i}_{\mathrm{contam}}$. 
To clarify our approach, we note that 
the right-hand side of Eqn.~(\ref{eqn:clcontam})
can be computed, with reasonable assumptions about the value of the 
cluster SN Ia rate. We can then compare this to the {\it observed}
value of the left-hand side of Eqn.~(\ref{eqn:clcontam}), 
and thereby infer
the contamination rate of the SN Ia sample due to finite 
redshift uncertainties.
For cuts on 
redshift consistency of 
$n_{\sigma} = 1.5$ 
and 
$n_{\sigma} = 2.5$, 
$N_{\mathrm{contam}}/N^{T}$ 
has the values 
$0.151~(r_{\mathrm{cl}}/\mathrm{SNu}r)^{-1}$ 
and
$0.276~(r_{\mathrm{cl}}/\mathrm{SNu}r)^{-1}$,
where $r_{\mathrm{cl}}$ is the assumed value of the cluster
SN rate, 
and 
$\mathrm{SNu}r = (10^{10} ~L^{r}_{\sun})^{-1} (100 ~\mathrm{yr})^{-1}$.

We first note that using values for 
$n_{\sigma} \approx 1.5-2.5$, we consistently
arrive at $N^{T} \approx 5$. 
By way of example, if we choose $n_{\sigma} = 2.5$ and 
$r_{\mathrm{cl}} = 0.6 ~\mathrm{SNu}r$, then we find
$N^{O}/N^{T} = 1.44$ (r.h.s. of Eqn.~(\ref{eqn:clcontam})). 
There are 7 SNe Ia with 
photometrically determined redshifts that are within
$\pm 2.5 \sigma$ of the cluster redshift 
(Table \ref{tab:snec4}), and so this implies a value of
$N^{T} = N^{O}/1.44 \approx 5$
(l.h.s. of Eqn.~(\ref{eqn:clcontam})).
Therefore, we apply a bias 
correction to our maxBCG SN sample by assuming that the 
number of photometric SNe in the sample
is 5, rather than the 7 listed in Table \ref{tab:snebcg}.
To estimate the systematic error on the size of the maxBCG SN sample,
we vary the assumed power-law SN rate model
according to the errors given in \citet{Dilday_08a}, and vary the
assumed cluster rate as $r_{\mathrm{cl}} = (0.6 \pm 0.2)~\mathrm{SNu}r$,
which results in an estimated systematic error of 
$\approx (-2-+3)\%$ on the maxBCG cluster SN rate 
measurements, due to employing photometric redshifts for 
a subset of the cluster SNe.

\subsection{Cluster Incompleteness and Contamination}
Here we discuss the effect of incompleteness and 
contamination in the galaxy cluster catalogs on
our cluster SN rate measurements, and on our comparisons of the 
cluster SN rate to the rate in early type galaxies in the field.
In what follows we assume that the SN rate in both galaxy clusters and 
the field is proportional to the host luminosity, with proportionality
constants $\alpha_{c}$ and $\alpha_{f}$ for clusters and the field, respectively.
We note that the C4 cluster catalog 
has $\approx 10\%$ incompleteness and
an $\approx 5\%$ contamination rate \citep{Miller_05}, 
and that the maxBCG catalog 
has $\approx 10\%$ incompleteness and
an $\approx 10\%$ contamination rate \citep{Koester_07b}.

{\bf Cluster Incompleteness}

With the ansatz that the cluster SN Ia rate is proportional to the
stellar luminosity, independent of cluster richness, cluster
incompleteness does not affect the derived value of the SN rate.
That is, although real clusters may fail to be included in the 
cluster catalog, the cluster SN sample will be reduced in
direct proportion to the luminosity of the missing clusters.
Cluster incompleteness, however, will affect the value of
the ratio of the rates for cluster vs.~field early-type galaxies, 
as SNe occurring in clusters may be erroneously associated with the field
SN sample.

In the presence of cluster incompleteness, the inferred
ratio of the SN Ia rate in clusters, $r_c$, to the SN rate in 
field early-type galaxies, $r_f$, will be,

\begin{eqnarray}
\label{eqn:systratioincomplete}
\frac{r_c}{r_f} & = & 
\left(\frac{N_c - N_{\delta-}}{L_c - L_{\delta-}}\right)
\left(\frac{L_f + L_{\delta-}}{N_f + N_{\delta-}}\right) \\
& \approx & \frac{\alpha_{c}}{\alpha_{f}}(1+(1-\alpha_{c}/\alpha_{f})(L_{\delta-}/L_c)(L_c/L_f)), 
\end{eqnarray}

\noindent where $N~(L)$ represents the true number of SNe (luminosity) in a given subset, 
and the $c$, $f$, and $\delta-$ subscripts denote clusters, the field, and the portion of 
the clusters missed due to incompleteness, respectively. The quantity $L_{\delta-}/L_c$ 
is the cluster incompleteness. 
To estimate the quantity $L_c/L_f$, we compute a cluster luminosity density
by dividing the summed $r$-band luminosity 
from the maxBCG cluster catalog (k-corrected from $^{0.25}i$ as described in \S \ref{sec:cllumcontent}), 
for clusters in the SN survey region, by the corresponding volume. For redshifts $ z \gtrsim 0.2$
this derived cluster luminosity density has the value 
$\approx 0.104 \times 10^{8}~L_{\sun}~\mathrm{Mpc}^{-3}~h$. 
In \citet{Dilday_08a} we showed that, using the luminosity functions of \citet{Blanton_03}, and 
an objective classification scheme for early-type galaxies, the $^{0.1}r$-band luminosity density, 
for early type galaxies is 
$0.994 \times 10^{8}~L_{\sun}~\mathrm{Mpc}^{-3}~h$. 
Applying a k-correction as
described in \S \ref{sec:cllumcontent}, this corresponds to a $r$-band luminosity density, for early-type
galaxies, of 
$1.24 \times 10^{8}~L_{\sun}~\mathrm{Mpc^{-3}}~h$. 
Therefore, we conclude that the ratio 
of the total $r$-band light in cluster galaxies 
vs.~field early-type galaxies, $L_c/L_f$, is $\approx 0.091$.
The fraction of cluster luminosity contained in early-type galaxies is $\approx 80\%$ \citep{Hansen_07}, 
and thus the ratio of the $r$-band light in cluster early-type galaxies 
vs.~field early-type galaxies, $(L_c/L_f)_{\mathrm{red}}$, is $\approx 0.073$.
In turn, this results in an estimated $\approx -(1-2)\%$ correction factor, for values of the ratio of the SN 
rates, $\alpha_{c}/\alpha_{f}$, of $\approx 2-3$.  

{\bf Cluster Contamination}

Contamination of the cluster catalogs will affect both the
value of the SN rate in clusters, and the ratio of the rate in clusters to the
rate in the field. 

The inferred cluster SN rate will be,
\begin{eqnarray}
\label{eqn:systratecontam}
r_c  & = & \frac{N_c + N_{\delta+}}{L_c + L_{\delta+}} \\
& \approx & \alpha_{c}~(1+(\alpha_{f}/\alpha_{c}-1)~L_{\delta+}/L_c), 
\end{eqnarray}

\noindent where $N$ and $L$ are defined as in equation \ref{eqn:systratioincomplete},
and $\delta+$ denotes field quantities that were erroneously identified with 
clusters. The quantity $L_{\delta+}/L_c$ represents the cluster contamination rate.
The estimated correction factor on the SN Ia cluster rate measurement due to cluster contamination is then
$\approx +3\%$ for C4 clusters, and $\approx +(5-7)\%$ for maxBCG clusters.

The inferred ratio of the SN rate will be,
\begin{eqnarray}
\label{eqn:systratiocontam}
\frac{r_c}{r_f} & = & 
\left(\frac{N_c + N_{\delta+}}{L_c + L_{\delta+}}\right)
\left(\frac{L_f - L_{\delta+}}{N_f - N_{\delta+}}\right) \\
& \approx & \frac{\alpha_{c}}{\alpha_{f}}(1+(\alpha_{f}/\alpha_{c}-1)~L_{\delta+}/L_c), 
\end{eqnarray}

\noindent with estimated correction factors on the cluster vs.~field SN rates of 
$\approx +3\%$ for C4 clusters, and $\approx +(5-7)\%$ for maxBCG clusters.

\subsection{Summary of SN Ia Rate Corrections}
\label{sec:clcorrsum}

To summarize our consideration of necessary corrections to our
cluster SN Ia rate measurements, we have considered
the effects of the lesser average intrinsic luminosity for SNe Ia in 
early-type galaxies as compared to a representative galaxy sample, 
contamination of the cluster SN Ia sample due to the 
use of SN photometric redshifts, and 
incompleteness (clusters are not identified by the cluster finding algorithms)
and contamination (objects that are not clusters are identified as clusters) 
of the galaxy cluster catalogs.
The lesser average intrinsic luminosity of SNe Ia in early-type 
galaxies does not affect the selection efficiency for the 
relatively low-redshift C4 SN sample, 
and has an $\approx 2\%$ effect on the maxBCG SN sample. 
This factor is accounted for through the efficiency function, 
$\epsilon(z)$, that is mentioned in \S \ref{sec:clresults}.
We estimate that $\approx 2$ of the 7 SNe with photometric
redshifts in the maxBCG SN Ia sample are not in fact associated 
with the cluster, and thus we will apply a correction of 
$f^{pz} \approx -2/27 = -7.4\%$ to the maxBCG cluster rate.
Cluster contamination results in correction factors 
of $f^{c} \approx +3\%$ 
for the C4 cluster SN Ia rate, 
and $f^{c} \approx +(5-7)\%$ for the maxBCG cluster SN Ia rate.
Cluster contamination also results in correction factors 
on the ratio of the SN Ia rate in early-type cluster galaxies to early-type field galaxies 
of $f^{c} \approx +3\%$ 
for the C4 clusters, and $f^{c} \approx +(5-7)\%$ for the maxBCG clusters.
Additionally, cluster incompleteness 
results in an $f^{I} \approx -(1-2)\%$ correction factor 
on the ratio of the SN Ia rate in early-type cluster galaxies to early-type field galaxies
for both the C4 and the maxBCG clusters. 
We note that in \citet{Dilday_10} it is shown that the contamination 
of the SDSS-II SN Survey {\it photometric} SN Ia sample 
by non-Ia SNe is not more than 
$\approx 3\%$. Photometric SNe Ia make up $\approx 50\%$ of the 
maxBCG cluster SN Ia sample, and so the overall contamination of the 
cluster SN Ia sample is $\lesssim 2\%$. Contamination of the
SN Ia sample due to lack of a spectroscopic identification is
likely to be correlated with the contamination due to the use 
of photometric redshifts discussed above, and so the size of the 
effect is expected to be less than this.

\section{Systematic Errors}
\label{sec:systerrors}

Here we discuss possible sources of systematic error on
our cluster SN Ia rate measurements. 
In the previous section we derived correction factors to 
account for incompleteness and contamination of the 
galaxy cluster catalogs. These correction factors will
each have an uncertainty due to sample variance in the 
galaxy clusters in the SN survey region.
We evaluate the uncertainty by assuming that each galaxy cluster 
in the sample will suffer from incompleteness or contamination 
with a probability given be the mean incompleteness and contamination
rates given above. The uncertainty on the correction factor 
is then derived from the variance of the corresponding binomial distribution,
with the number of events equal to the number of galaxy clusters 
considered. 
For the C4 clusters (71 clusters), this 
results in a 
$\pm 3.2 \%$ uncertainty 
on the correction for 
cluster incompleteness 
and a $\pm 2.5 \%$ uncertainty 
on the correction for 
cluster contamination.
For the maxBCG clusters (492 clusters), this 
results in a 
$-1.2\%$ - $+1.0\% $ uncertainty 
on the correction for 
both cluster incompleteness and
cluster contamination. 
\citet{Dilday_10} discuss the systematic uncertainty on the 
SN Ia rate for the full SDSS-II SN Survey sample, and show that 
assuming a value for the mean extinction in $V$-band,
$\langle A_V \rangle =0.45$, as opposed to the fiducial value of 
$\langle A_V \rangle = 0.35$, has a large systematic
effect on the derived SN Ia rates for SNe Ia at $z \gtrsim 0.2$.
For the present sample of SNe Ia in galaxy clusters, assuming a larger than
fiducial value for the mean dust extinction is not appropriate, and
so we do not include this systematic effect on the
cluster SN Ia rate measurements.
As noted in 
\S \ref{sec:clust_uncert}, any uncertainty on the 
cluster luminosities will also contribute to the 
total uncertainty on the cluster SN Ia rate.

\section{SDSS SN Results}
\label{sec:clresults}

\subsection{The C4 cluster rate}
As discussed above, there are \nsnecf~type-Ia SNe 
in C4 clusters that satisfy the SN selection criteria.
The total $r$-band luminosity in the SN survey region, 
after correcting for the 
faint end of the LF, is 
$4.99 \times 10^{13} ~\mathrm{L}^{r}_{\sun}$ $h^{-2}$, 
and the mean redshift of the C4 clusters is
$\langle z \rangle = 0.0786$.
The observing time for the \sns~was 
89, 90, and 90 days, 
for the 2005, 2006, and 2007
observing seasons, respectively.
The SN selection efficiency is approximately constant over the range of the C4
clusters, with the values 0.77, 0.73, and 0.72 
for the 2005, 2006, and 2007
observing seasons, respectively.
Using these values, the SN rate in C4 clusters is
\begin{equation}
r = \frac{N (1+\langle z \rangle)g^{c}}{\Sigma L_{r} ~\Sigma_i(\epsilon ~T)} 
= {\cfsnur}_{-\cfsnurlo-\cfsnurlosyst}^{+\cfsnurhi+\cfsnurhisyst} ~\mathrm{SNu}r ~h^{2}
\end{equation}

\noindent where $i$ denotes each observing season,
$T$ is the survey observation time, 
$\epsilon$ is the SN selection efficiency,
$N$ is the number of SNe Ia,
$\Sigma L_{r}$ is the total cluster luminosity in the $r$ band,   
and $\mathrm{SNu}r = (10^{10} ~L^{r}_{\sun})^{-1} (100 ~\mathrm{yr})^{-1}$.
The factor $g^{c} = 1 + f^{c} = 1.03$, where $f^{c}$ is the correction 
due to contamination of the cluster catalog, discussed in 
\S \ref{sec:clcorrections}.
The errors quoted are the 1-sigma statistical and systematic errors, 
respectively.
We have used the mean survey efficiency, determined from 
Monte Carlo simulations, as both the positions of the
clusters and the Monte Carlo SNe are
effectively uniform random samplings of the 
survey area.
\citet{Jorgensen_97} (clusters) and \citet{Padmanabhan_04} (field)
give the average stellar mass to luminosity ratio ($r$-band) for early-type galaxies as
$\approx 3$, which is in good agreement with the M/L conversion 
employed by \citet{Sharon_07a}. Using this assumption, 
the Type Ia SN rate per unit
luminosity quoted above is equal to ${\cfsnum}^{+\cfsnumhi+\cfsnumhisyst}_{-\cfsnumlo-\cfsnumlosyst} ~\mathrm{SNuM} ~h^{2}$, with 
$\mathrm{SNuM} = (10^{10} ~M_{\sun})^{-1} (100 ~\mathrm{yr})^{-1}$. 
Using the same 
average conversion factor from $r$ to $B$ as \citet{Sharon_07a}, 
this corresponds to 
${\cfsnub}^{+\cfsnubhi+\cfsnubhisyst}_{-\cfsnublo-\cfsnublosyst} ~\mathrm{SNu}B ~h^{2}$.

\subsection{The maxBCG cluster rate}

\begin{figure} [t] 
\begin{center}
\includegraphics[width=5.75in]{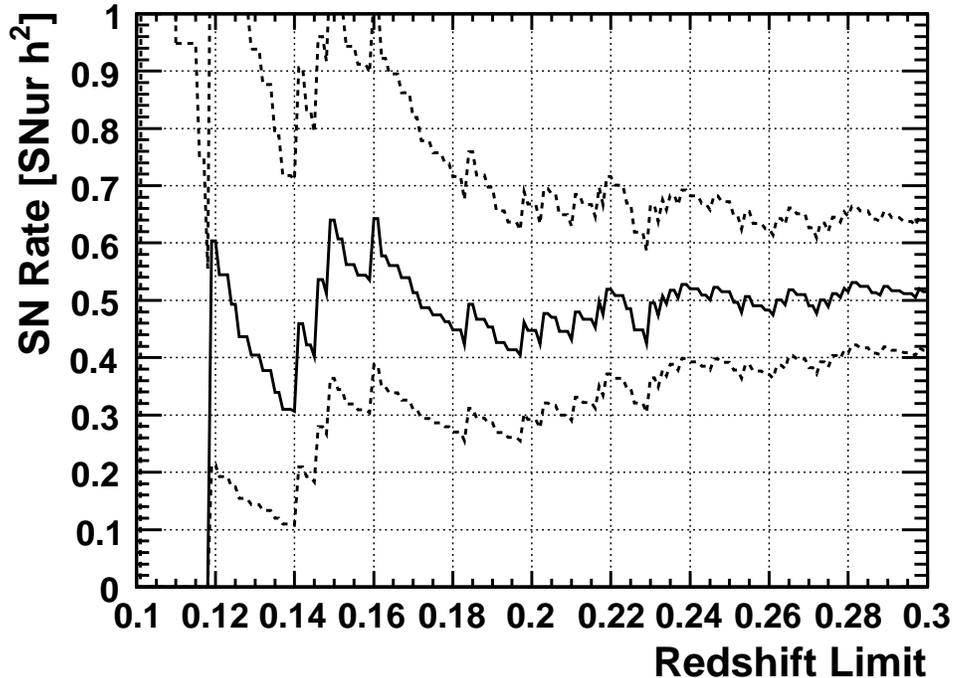}
\end{center}
\caption{
maxBCG cluster SN rate as a function of upper limit on the redshift 
range. 
The dashed lines represent the 1-sigma upper and lower 
limit of the SN rate.
}
\label{fig:figbcgratez}
\end{figure}

\begin{figure} [t] 
\begin{center}
\includegraphics[width=5.75in]{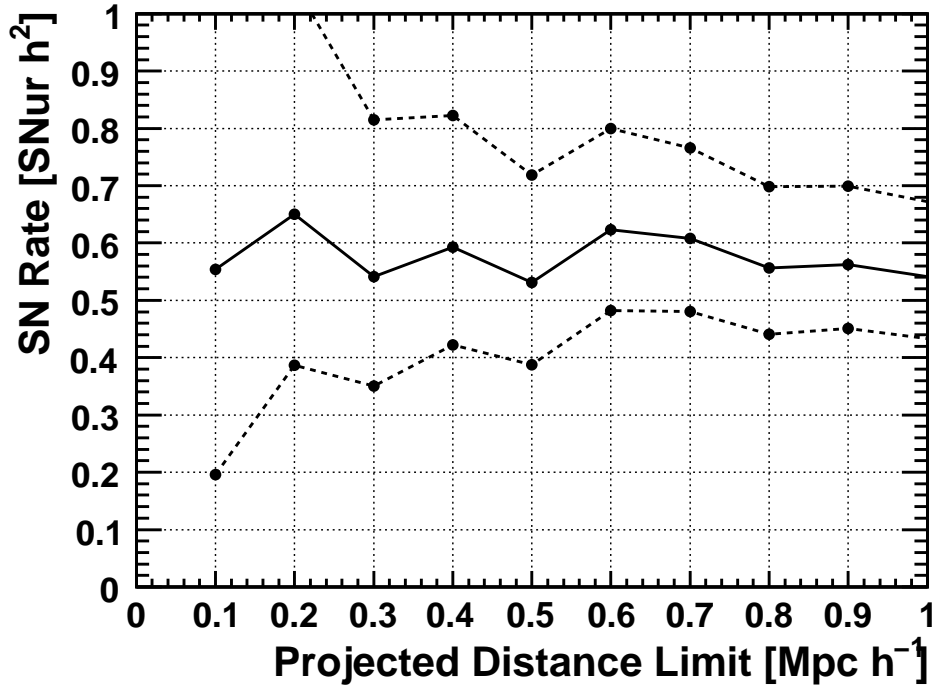}
\end{center}
\caption{
maxBCG cluster SN rate as a function of limit on the projected distance.
The dashed lines represent the 1-sigma upper and lower 
limit of the SN rate.
 }
\label{fig:figbcgratedist}
\end{figure}

\begin{figure} [t] 
\begin{center}
\includegraphics[width=5.75in]{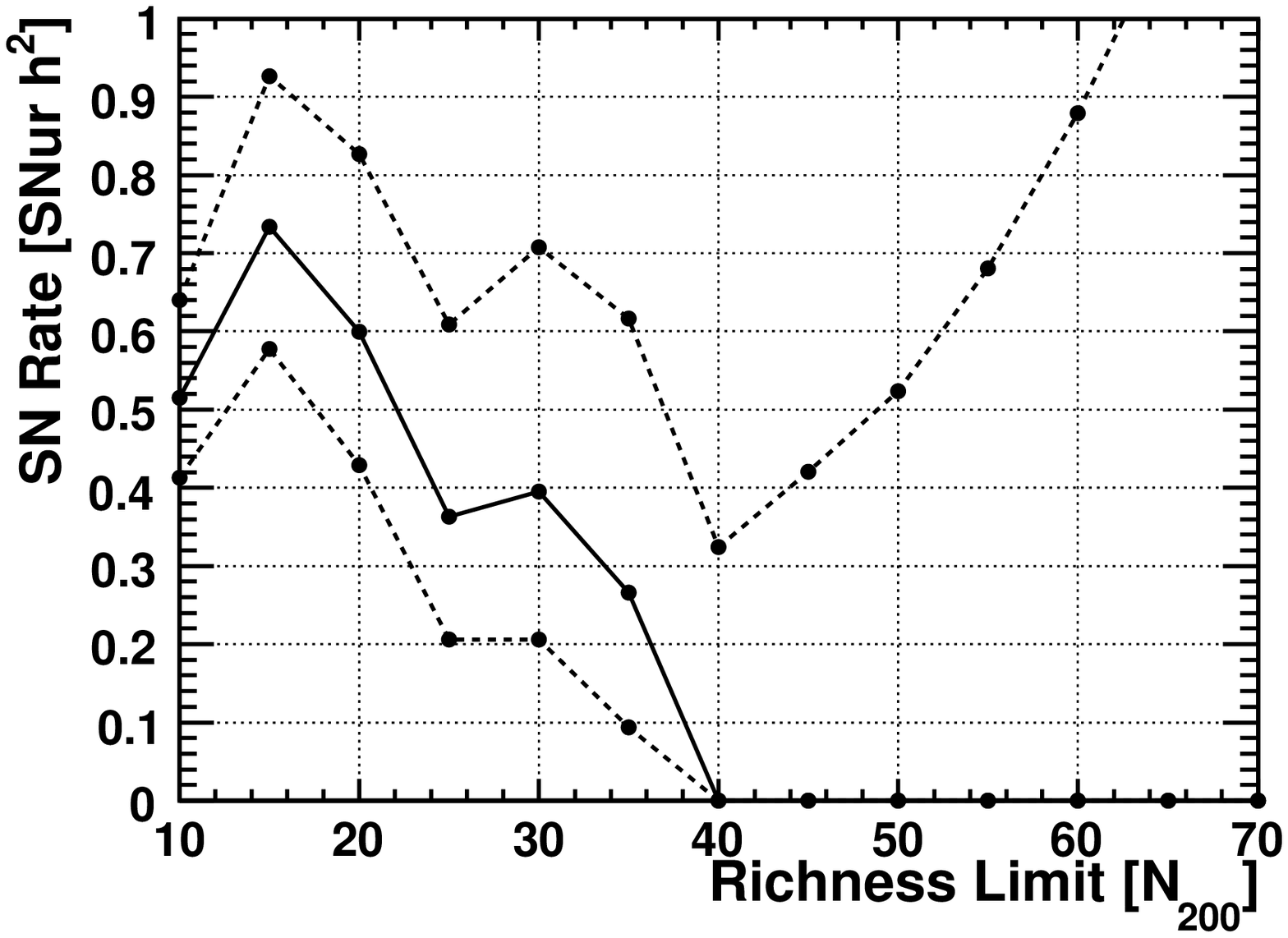}
\end{center}
\caption{
maxBCG cluster SN rate as a function of lower limit on the cluster
richness measure, $N_{200}$.
The dashed lines represent the 1-sigma upper and lower 
limit of the SN rate.
}
\label{fig:figbcgraten200}
\end{figure}

As discussed in \S \ref{sec:clsample},
there are \nsnebcg~SNe in maxBCG clusters 
from \sns~that satisfy
the selection criteria. 
As the \sns~SN discovery efficiency is not well approximated as constant 
over the redshift range of the maxBCG catalog, 
in determining the SN rate in maxBCG clusters
we use the more formal definition of 
the SN rate per unit luminosity, $r_{L}$,

\begin{equation}
r_{L} = \frac{N g^{pz} g^{c}}{\widetilde{\epsilon T \Sigma L}}
\end{equation}

with

\begin{equation}
\widetilde{\epsilon T \Sigma L} = T_{\earth} ~\int_{z_{\mathrm{min}}}^{z_{\mathrm{max}}}~dz ~\frac{\Sigma L(z) ~\epsilon(z)}{1+z},
\end{equation}

\noindent where $T_{\earth}$ is the earth frame observation time,
$\Sigma L(z)$ is the total cluster luminosity as a function of
redshift, and $\epsilon(z)$ is the SN discovery efficiency.
The factors $g^{pz}$ and $g^{c}$ represent the corrections
due to use of SN photometric redshifts and contamination of the 
cluster catalog, respectively (\S \ref{sec:clcorrections}).
These factors have the values 
$ g^{pz} = 1 + f^{pz} = 0.926$ and 
$ g^{c} = 1 + f^{c} = 1.06$.
For the maxBCG catalog $z_{\mathrm{min}}$ is fixed to 0.1. 
In Figure \ref{fig:figbcgratez} we show the value of the maxBCG cluster
SN rate as a function of $z_{\mathrm{max}}$. This figure shows that
the derived SN rate is not strongly sensitive to the upper limit
on the cluster sample. If the upper limit is chosen as
the upper limit of the maxBCG catalog, $z = 0.3$, then
we have $N = \nsnebcg$ 
and 
$\widetilde{\epsilon T \Sigma L} = \bcgeplt $
$\mathrm{yr} ~L^{r}_{\sun} ~h^{-2}$, including the correction for the 
faint end of the LF. The derived value 
of the SN rate is thus
$r_{L} = {\bcgsnur}^{+\bcgsnurhi+\bcgsnurhisyst}_{-\bcgsnurlo-\bcgsnurlosyst}~\mathrm{SNu}r ~h^{2}$
$(= {\bcgsnub}^{+\bcgsnubhi+\bcgsnubhisyst}_{-\bcgsnublo-\bcgsnublosyst}~\mathrm{SNu}B ~h^{2}$
$ = {\bcgsnum}^{+\bcgsnumhi+\bcgsnumhisyst}_{-\bcgsnumlo-\bcgsnumlosyst}~\mathrm{SNu}M ~h^{2})$.
Figure \ref{fig:figbcgratedist} shows the cluster SN rate as a function
of the limit on the projected distance from the center of the cluster,
in units of $\mathrm{Mpc} ~h^{-1}$. The fraction of red, early-type 
galaxies in clusters is larger at small separations from the cluster
center, and in this sense the SN rate at smaller separations
is a more reliable probe of the component of the SN rate that originates from 
an old stellar population. Furthermore, the extent of a 
cluster is not an unambiguously defined quantity, and SNe at smaller
separations are more robustly associated with the cluster. 
Figure \ref{fig:figbcgratedist} shows that the derived SN rate is 
not strongly dependent on the limiting projected radius that we
use to define cluster membership.
Figure \ref{fig:figbcgraten200} shows the SN rate 
as a function of the lower limit on the cluster richness
measure, $N_{200}$.

\begin{figure} [t] 
\begin{center}
\includegraphics[width=5.75in]{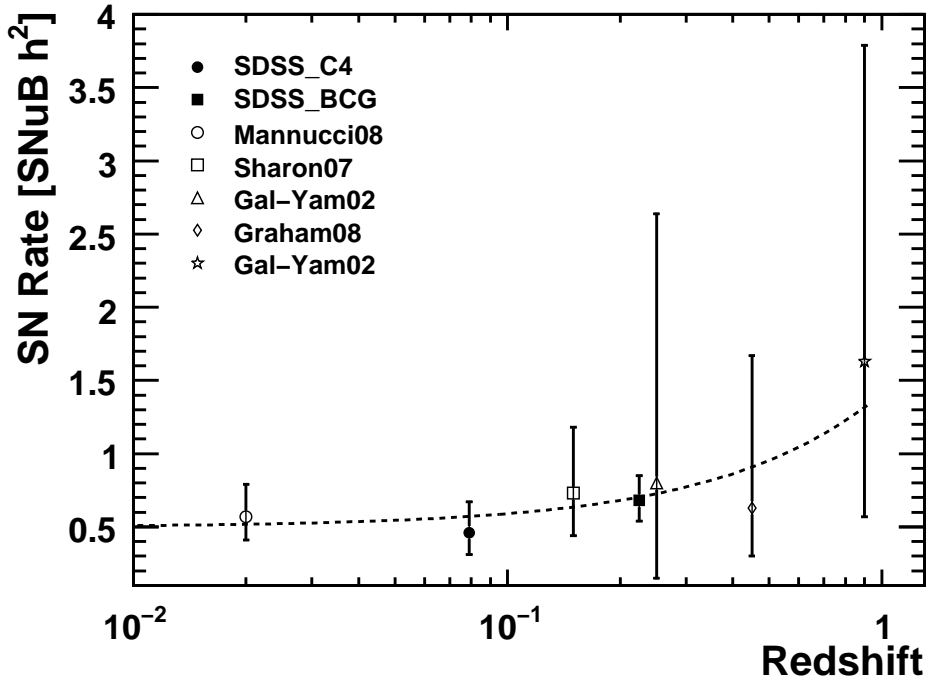}
\end{center}
\caption{
Cluster SN Ia rate vs.~redshift. The dashed line shows the
best fit to a linear model of the SN rate as a function of redshift.
 }
\label{fig:fclratez}
\end{figure}

The \sns~cluster rate results, along with the previous 
measurements listed in Table \ref{tab:clraterev}, 
in units of $\mathrm{SNu}B ~h^{2}$, are shown in 
Figure \ref{fig:fclratez}. 
A fit of the data to a linear 
model of the cluster SN Ia rate as a function of redshift, 
$r = A + B z$, gives best fit values of 
$A = 0.49 \pm 0.14$ 
$\mathrm{SNu}B$ $h^{2}$
and 
$B = 0.81^{+0.82}_{-0.80}$ 
($\chi^2/NDF = 0.88/5$). 
A fit of the data to a constant
model of the cluster SN Ia rate as a function of redshift, 
$r = A$, gives a best fit value of 
$A = 0.61^{+0.95}_{-0.89} $ 
$\mathrm{SNu}B$ $h^{2}$
($\chi^2/NDF = 1.9/6$). 

\subsection{Cluster SN Rate vs.~Field SN Rate}
In \citet{Dilday_08a}, the SN rate in low-redshift 
early-type galaxies was estimated from the first year data of the 
\sns~as
$\approx 0.17_{-0.04}^{+0.06}$ $ \mathrm{SNu}r ~h^{2}$. 
One SN in the 
sample described in \citet{Dilday_08a} is a cluster SN, and hence the 
early-type {\it field} rate is 
$\approx 0.16_{-0.04}^{+0.06}$ $ \mathrm{SNu}r ~h^{2}$. 
Using the same color cut ($u-r = 2.4$) to differentiate early and late
type galaxies as \citet{Dilday_08a}, we find that 6/9 of the C4 cluster SNe
reside in early-type galaxies, while 20/27 maxBCG cluster SNe reside in 
early-type galaxies (18/25 after bias correction to account for SN photometric redshifts).
According to \citet{Hansen_07}, the fraction of light in early-type galaxies 
for the maxBCG clusters is $\approx 80\%$.
Therefore the cluster SN rate for early-type galaxies is 
$r_{L} = {\cfsnurearly}_{-\cfsnurloearly-\cfsnurlosystearly}^{+\cfsnurhiearly+\cfsnurhisystearly} ~\mathrm{SNu}r ~h^{2}$
and
$r_{L} = {\bcgsnurearly}^{+\bcgsnurhiearly+\bcgsnurhisystearly}_{-\bcgsnurloearly-\bcgsnurlosystearly}~\mathrm{SNu}r ~h^{2}$
for C4 and maxBCG clusters, respectively.
As discussed in \S \ref{sec:clcorrections}, 
the ratio of the cluster early-type to field early-type 
rates must be multiplied by a factor $g^{I} = 1 - f^{I} = 0.985$
to account for incompleteness of the galaxy cluster catalogs.
The C4 and maxBCG cluster SN Ia rates 
are seen to be larger by factors of 
${\cfrat}^{+\cfrathi+\cfratsysthi}_{-\cfratlo-\cfratsystlo}$
and
${\bcgrat}^{+\bcgrathi+\bcgratsysthi}_{-\bcgratlo-\bcgratsystlo}$
compared 
to the field SN rate for early-type galaxies, respectively.
This is broad agreement with the enhancement 
of the Type Ia SN rate in early-type galaxies in 
galaxy clusters reported by \citet{Mannucci_08}.

\subsection{The SN Ia Rate in Cluster BCGs}
Of the 32 SNe in our cluster sample 
(after accounting for duplicates and for contamination due to use
of SN photometric redshifts), 2 are in the brightest cluster galaxy (BCG) 
of the cluster. These are SNe 12979 and 13905.
Another, SN 18890 appears to be in the BCG, although it 
has a somewhat ambiguous host.
All 3 of these SNe are in C4 clusters, while 1 is in a maxBCG cluster.
The cluster BCGs contain $\approx 6\%$ of the total $r$-band cluster luminosity,
and so the SN Ia rate, in the cluster BCGs, is
$r_{L} = {\cfsnurbcg}^{+\cfsnurhibcg+\cfsnurhisystbcg}_{-\cfsnurlobcg-\cfsnurlosystbcg}~\mathrm{SNu}r ~h^{2}$
for the C4 clusters 
and
$r_{L} = {\bcgsnurbcg}^{+\bcgsnurhibcg+\bcgsnurhisystbcg}_{-\bcgsnurlobcg-\bcgsnurlosystbcg}~\mathrm{SNu}r ~h^{2}$
for the maxBCG clusters.

\subsection{Hostless SNe}
For the SNe Ia in our cluster sample, all but 2 are visually associated with a host galaxy.
The 2 SNe Ia with no host evident in the images are SN 13073 and SN 19001.
Additionally, one SN, SN 1782, lies at a distance $\approx 25 ~\mathrm{kpc} ~h^{-1}$ from its presumed host galaxy, and might 
reasonably be considered to be hostless. 
Assuming SN 1782 is in fact a hostless SN, we can 
constrain the fraction of hostless SNe in galaxy clusters to be 
$\approx 9.4\%$, with 1-sigma confidence interval $(4.3-17.7)\%$. For comparison, 
\citet{Galyam_03} reported 2 hostless SNe from a sample of 7 for an estimated 
fraction of hostless SNe of $\approx 30\%$.
We note, however, that 
the classification of these 2 SNe as 
hostless by \citet{Galyam_03} relied, in part, on 
the observation of velocity offsets between the SNe and its potential host of 
$\approx 750-2000~\mathrm{km/s}$. In principle a more thorough analysis of 
SN and galaxy spectra for SNe from the \sns~may reveal additional hostless SNe.
The clusters considered by \citet{Galyam_03} were generally more massive than 
the clusters considered here, and an additionally possibility is that the 
rate of hostless SNe Ia in galaxy clusters is larger for more massive clusters.

\subsection{Radial Distribution of Cluster SNe}
\label{sec:clraddist}
\citet{Forster_08} have studied the radial distribution of SNe in 
early-type galaxies. The sample of SNe includes an unspecified number of 
SNe from the \sns. 
Here we consider the radial distribution for SNe for field and cluster ellipticals.

\begin{figure} [t] 
\begin{center}
\plottwo{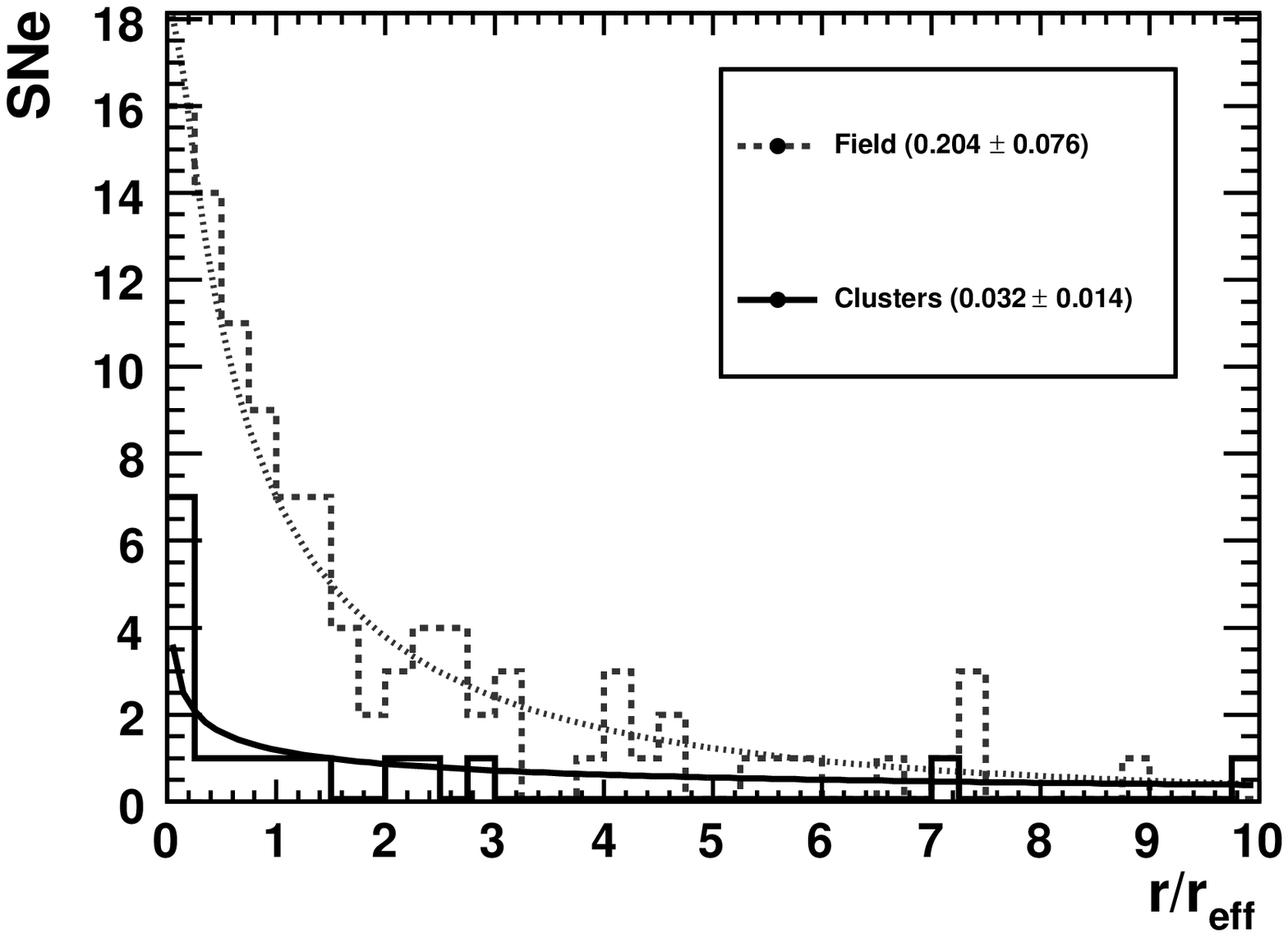}{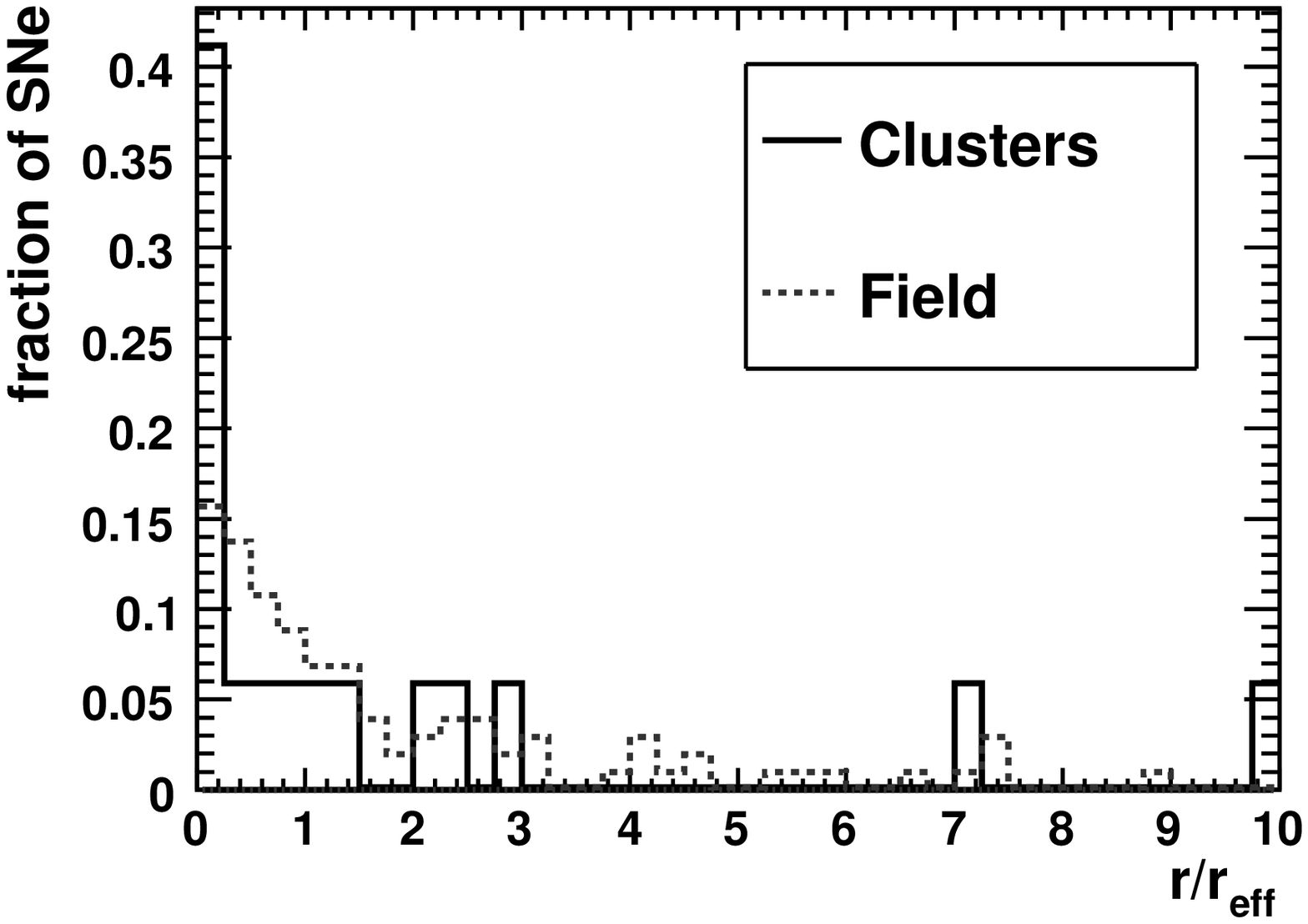}
\end{center}
\caption{
Radial distribution of SNe in early-type host galaxies.
 }
\label{fig:figcandhostdist2}
\end{figure}

To construct the sample of early-type galaxies, we employ the following procedure.
We first assign a host galaxy to each SNe by locating the nearest host from the SDSS database, 
in units of the isophotal radius of the host galaxy in $r$-band. The algorithm is described
in detail in \citet{Dilday_08a}. 
Early-type galaxies are defined as those host galaxies that 
satisfy the following criteria:

\begin{itemize}
\item $u-r > 2.4$ 
\item $r < 21.5$ 
\item $\Delta r < 0.05$ 
\end{itemize}

\noindent where $u$ and $r$ are the SDSS model magnitudes and
$\Delta r$ is the error on the $r$-band magnitude.
It is a well established result that the $(u-r)$ color for SDSS galaxies
is bimodal \citep{Strateva_01}, with early-type galaxies generally having $(u-r) > 2.2$.
As discussed in \citet{Dilday_08a}, studies of the observed $u-r$ distribution for
galaxies from the photometric redshift catalog of \citet{Oyaizu_07a} suggest that 
$u-r = 2.4$ may provide a more robust separation.
The requirement that $r < 21.5$ is imposed as the separation of stars and 
galaxies is fairly robust to this limit. The requirement that $\Delta r < 0.05$
is imposed to remove outlying, poorly measured galaxies.  
The distributions of 
the distance of each SN from its host galaxy, 
for field and cluster early-type hosts, 
is shown in 
Figure \ref{fig:figcandhostdist2}.
For field early-type galaxies, a fit of the data to a Sersic model of the 
luminosity distribution, 
$dN/dr = A ~\rho ~e^{-\gamma \rho^{-\lambda}}$, where $\rho$ 
denotes the distance of the SN in units of 
the \deV ~radius of the host galaxy, gives a value for $\lambda$ 
of $0.20 \pm 0.08$, which is consistent 
with a \deV ~profile ($\lambda = 0.25$).
Since the distribution of light in early-type galaxies is known to follow a 
\deV ~profile, the result of the fit confirms the results of 
\citet{Forster_08}  that the SN rate in field
ellipticals is well represented by a constant rate per unit luminosity. 
The radial distribution for SNe in cluster early-type galaxies 
shows an an enhancement at small radial separations in comparison to 
a \deV~profile.
An enhancement of the 
SN Ia rate in regions that have undergone recent star formation 
has been reported by several authors (see, e.g., the review 
by \citet{Mannucci_09}), and
our observed enhancement at small radial separations is possible evidence for a 
component of the cluster SN Ia rate 
that tracks residual star formation activity in cluster early-type galaxies.
Such an enhancement is in qualitative agreement with the larger 
SN Ia rate in cluster ellipticals, compared with field ellipticals, 
mentioned above and by \citet{Mannucci_08}, as well as with recent 
independent evidence for some recent star-formation activity in 
early-type galaxies (e.g, \citet{Kaviraj_07}). 
Additionally, \citet{DellaValle_05} 
and \citet{Graham_10} report an enhancement of the 
SN Ia rate in radio-loud early-type galaxies, and attribute this to 
mergers that not only power the radio emission, but also provide 
a young stellar population that can account for the enhanced SN Ia rate. 
While we do not consider the radio properties of the early-type galaxies in 
our sample, it is possible that our results are related to the results 
of \citet{DellaValle_05} and \citet{Graham_10}.
While the best fit radial profile 
for cluster early type galaxies does not match a 
\deV~profile, a KS test on the radial distributions of SNe Ia in cluster and 
field ellipticals results in an $\approx 30\%$ 
probability that the data are drawn from the same underlying distribution.
As discussed in~\citet{Dilday_08a}, all SN selection efficiency calculations 
are based ultimately on artificial SNe Ia inserted directly into the survey search imaging data.
Analysis of these artificial SNe Ia from the three observing seasons of the \sns~does not 
show evidence for significant loss of efficiency near the cores of galaxies. 
While the artificial SNe were inserted into random galaxies that on the average may be less luminous in the cores than
typical cluster galaxies, any additional inefficiency would only increase the observed 
enhancement of the SN Ia rate in the cores of cluster early-type galaxies.

\section{Conclusions}
\label{sec:clconclude}

We have presented measurements of the Type Ia SN rate in galaxy clusters 
over the redshift range $0.03 < z < 0.30$. These measurements are based on 
\nsnetot~SNe Ia (32 after applying a bias correction) from the \sns~and 
represent a significant statistical 
contribution to the study of the Type Ia SN rate in galaxy clusters. Our 
results on the Type Ia SN rate are consistent with previously published 
measurements, both in the local universe, and at redshift $\approx 0.15-0.25$.
In turn, the local and low-redshift SN cluster rates are consistent with the 
SN Ia rate at $z \approx 0.45$ and at $z \approx 0.9$. 
The current cluster SN Ia rate measurements do not show evidence for
a rapid increase in the SN rate, as a function of lookback time, as 
has been well-established for the volumetric SN Ia rate. 
It should be emphasized, however, that the existing cluster SN rate measurements 
are based on small samples and the measurements do not rule out
a redshift dependence to the cluster SN rate. A fit of the cluster SN Ia
rate measurements to a linear dependence on redshift results in 
a best fit slope of 
$(0.91^{+0.85}_{-0.81})~\mathrm{SNu}B$ $h^{2}$ per unit redshift.
We find a ratio of the SN Ia rate in cluster early-type galaxies
to that of the SN Ia rate in field early-type galaxies of
${\cfrat}^{+\cfrathi+\cfratsysthi}_{-\cfratlo-\cfratsystlo}$
and
${\bcgrat}^{+\bcgrathi+\bcgratsysthi}_{-\bcgratlo-\bcgratsystlo}$,
for C4 and maxBCG clusters, respectively.

We find that at most 3 of the SNe Ia in our sample are hostless, intra-cluster
SNe, which is 
significantly less than 
the $30\%$ hostless fraction
estimated from previous cluster SN studies.

We have presented the first study of the radial distribution of 
SNe Ia in cluster early-type galaxies. 
The radial distribution for SNe in cluster early-type 
galaxies shows an enhancement at small radial 
separations in comparison to the radial distribution in 
field early-type galaxies, which are well described by 
a \deV~profile.
This enhancement could be attributable 
to residual star formation in cluster early-type galaxies, 
which could explain the higher SN Ia rate observed in 
cluster early-type galaxies in comparison to early-type galaxies in the 
field.

\clearpage
\begin{deluxetable}{lccrrllcrllll}
\tablecolumns{13}
\tabletypesize{\scriptsize}
\singlespace
\rotate
\tablewidth{0pc}
\tablecaption{SNe in C4 Clusters
\label{tab:snec4}
}
\tablehead{
\colhead{SDSS Id} &   
\colhead{IAU} &   
\colhead{SN\tablenotemark{a}} &    
\colhead{Cluster} &    
\colhead{Cluster} &    
\colhead{Cluster} & 
\colhead{SN} & 
\colhead{Projected} & 
\colhead{$N_{\mathrm{gals}}$} &
\colhead{$u$\tablenotemark{d}} &
\colhead{$r$\tablenotemark{d}} &
\colhead{$u-r$\tablenotemark{d}} &
\colhead{$r_{\mathrm{abs}}$ \tablenotemark{c}} \\

\colhead{} &
\colhead{Name} &   
\colhead{Type} &
\colhead{Ra (J2000)} &    
\colhead{Dec (J2000)} &    
\colhead{Redshift} &
\colhead{Redshift} &
\colhead{Dist. (Mpc $h^{-1}$)} &

}

\startdata
6295  &  2005js    &  Ia  & 23.72890 &  -0.65160  &  0.0814  &  0.0786   &  0.2804  &  38  &  $19.268 \pm 0.052$  &  $16.301 \pm 0.004$  &  $2.967 $  &  $-21.663 $ \\
\hline
12979  &  2006gf   &  Ia  & 11.56640 &  +0.00200  &  0.1140  &  0.1153   &  0.1831  &  34  &  $18.404 \pm 0.052$  &  $15.438 \pm 0.003$  &  $2.966 $  &  $-23.469 $ \\
13905  &  \nodata  &  Ia-photo+z  & 11.54900 &  +0.34870  &  0.1153  &  0.1151   &  0.7008  &  5  &  $19.112 \pm 0.058$  &  $16.224 \pm 0.004$  &  $2.888 $  &  $-22.679 $ \\
14279  &  2006hx   &  Ia  & 18.78670 &  +0.29720  &  0.0450  &  0.0444   &  0.6849  &  128  &  $18.179 \pm 0.027$  &  $15.834 \pm 0.003$  &  $2.345 $  &  $-20.782 $ \\
16215  &  2006ne   &  Ia  & 18.78670 &  +0.29720  &  0.0450  &  0.0455   &  0.8912  &  128  &  $19.532 \pm 0.088$  &  $16.766 \pm 0.007$  &  $2.766 $  &  $-19.907 $ \\
16280\tablenotemark{b}  &  2006nz   &  Ia  & 14.27050 &  -0.91880  &  0.0443  &  0.0370   &  0.7514  &  13  &  $19.347 \pm 0.045$  &  $16.404 \pm 0.004$  &  $2.943 $  &  $-19.793 $ \\
16280\tablenotemark{b}  &  2006nz   &  Ia  & 14.06720 &  -1.25510  &  0.0442  &  0.0370   &  0.1348  &  65  &  $19.347 \pm 0.045$  &  $16.404 \pm 0.004$  &  $2.943 $  &  $-19.793 $ \\
\hline
18375  &  2007lg   &  Ia  & 11.56640 &  +0.00200  &  0.1140  &  0.1169   &  0.2683  &  34  &  $20.365 \pm 0.081$  &  $18.167 \pm 0.008$  &  $2.198 $  &  $-20.775 $ \\
18890  &  2007mm   &  Ia  & 16.43170 &  -0.84980  &  0.0669  &  0.0654   &  0.2968  &  5  &  $20.849 \pm 0.159$  &  $17.888 \pm 0.008$  &  $2.961 $  &  $-19.636 $ \\
19155  &  2007mn   &  Ia  & 31.01580 &  +0.25880  &  0.0770  &  0.0760   &  0.9720  &  8  &  $19.298 \pm 0.070$  &  $16.961 \pm 0.006$  &  $2.337 $  &  $-20.922 $ \\
19968  &  2007ol   &  Ia  & 24.34470 &  -0.44720  &  0.0558  &  0.0551   &  0.3693  &  19  &  $18.800 \pm 0.037$  &  $16.290 \pm 0.004$  &  $2.510 $  &  $-20.829 $ \\
\enddata
\tablecomments{SDSS Id denotes internal candidate designation.
The horizontal lines delineate SNe discovered during the 3 distinct observing seasons of the \sns.
}
\tablenotetext{a}{``Ia'' refers to spectroscopically identified SN; 
see \citet{Zheng_08}. ``Ia-photo+z'' refers to photometrically 
identified SN {\it with}
a spectroscopically measured host galaxy redshift; 
see \citet{Dilday_10}.}
\tablenotetext{b}{SN does not satisfy the color-typing criteria. 
See \citet{Dilday_10}.}
\tablenotetext{b}{Value has been k-corrected to the rest frame of the galaxy; see \S \ref{sec:kcors}}
\tablenotetext{c}{Observer frame magnitudes for the SN host galaxy.}
\end{deluxetable}

\begin{deluxetable}{lccrrllcclllll}
\tablecolumns{14}
\tabletypesize{\scriptsize}
\singlespace
\rotate
\tablewidth{0pc} 
\tablecaption{SNe in maxBCG Clusters.
\label{tab:snebcg}
}
\tablehead{
\colhead{SDSS Id} &
\colhead{IAU} &
\colhead{SN\tablenotemark{a}} &    
\colhead{Cluster} &    
\colhead{Cluster} &    
\colhead{Cluster} & 
\colhead{SN} & 
\colhead{SN \tablenotemark{c} } &
\colhead{Projected} & 
\colhead{$N_{200}$} &
\colhead{$u$ \tablenotemark{d}} &
\colhead{$r$ \tablenotemark{d}} &
\colhead{$u-r$ \tablenotemark{d}} &
\colhead{$r_{\mathrm{abs}}$ \tablenotemark{b}} \\

\colhead{} &
\colhead{Name} &
\colhead{Type} &
\colhead{Ra (J2000)} &    
\colhead{Dec (J2000)} &    
\colhead{Redshift} &
\colhead{Redshift} &
\colhead{Redshift Err.} &
\colhead{Dist. (Mpc $h^{-1}$)} &
\colhead{} &

}

\startdata
1008  &  2005il  &  Ia-photo+z  & 28.17341 &  +1.13440  &  0.2025  &  0.2251  &  \nodata  &  0.8972  &  19  &  $23.036 \pm 0.855$  &  $19.707 \pm 0.030$  &  $3.329 $  &  $-20.963 $ \\
1740  &  \nodata  &  Ia-photo+z  & 5.34753 &  -0.82581  &  0.1837  &  0.1661  &  \nodata  &  0.6152  &  25  &  $21.553 \pm 0.235$  &  $18.485 \pm 0.012$  &  $3.068 $  &  $-21.362 $ \\
1782  &  \nodata  &  Ia-photo  & 337.85896 &  +0.29531  &  0.2781  &  0.2382  &  0.0189  &  0.5092  &  12  &  $22.360 \pm 0.731$  &  $17.682 \pm 0.008$  &  $4.678 $  &  $-23.146 $ \\
5549  &  2005hx  &  Ia  & 3.21852 &  +0.22231  &  0.1405  &  0.1198  &  \nodata  &  0.2569  &  16  &  $21.847 \pm 0.347$  &  $20.832 \pm 0.076$  &  $1.015 $  &  $-18.172 $ \\
5717  &  2005ia  &  Ia  & 17.95342 &  -0.01817  &  0.2646  &  0.2506  &  \nodata  &  0.6039  &  16  &  $22.807 \pm 0.456$  &  $21.712 \pm 0.098$  &  $1.095 $  &  $-19.258 $ \\
8280  &  \nodata  &  Ia-photo+z  & 8.59684 &  +0.85725  &  0.1972  &  0.1838  &  \nodata  &  0.5378  &  32  &  $19.844 \pm 0.122$  &  $18.328 \pm 0.020$  &  $1.516 $  &  $-21.789 $ \\
\hline
12979  &  2006gf  &  Ia  & 11.60084 &  +0.00238  &  0.1189  &  0.1153  &  \nodata  &  0.0069  &  23  &  $18.404 \pm 0.052$  &  $15.438 \pm 0.003$  &  $2.966 $  &  $-23.469 $ \\
13073  &  \nodata  &  Ia-photo  & 335.99616 &  +0.10358  &  0.2889  &  0.3246  &  0.0336  &  0.5282  &  15  &  \nodata  &  \nodata  &  \nodata  &  \nodata  \\
13655  &  2006hs  &  Ia  & 39.03651 &  -1.00655  &  0.2727  &  0.2512  &  \nodata  &  0.2146  &  13  &  $21.435 \pm 0.264$  &  $18.647 \pm 0.014$  &  $2.788 $  &  $-22.330 $ \\
14340  &  \nodata  &  Ia-photo+z  & 345.81978 &  -0.85395  &  0.2754  &  0.2762  &  \nodata  &  0.0741  &  37  &  $23.409 \pm 1.323$  &  $18.465 \pm 0.012$  &  $4.944 $  &  $-22.783 $ \\
15201  &  2006ks  &  Ia  & 337.53267 &  -0.00373  &  0.2188  &  0.2073  &  \nodata  &  0.1352  &  32  &  $21.382 \pm 0.369$  &  $18.175 \pm 0.013$  &  $3.207 $  &  $-22.268 $ \\
15823  &  \nodata  &  Ia-photo+z  & 314.20449 &  +0.25007  &  0.2296  &  0.2142  &  \nodata  &  0.6484  &  22  &  $21.944 \pm 0.439$  &  $19.248 \pm 0.023$  &  $2.696 $  &  $-21.285 $ \\
16021  &  2006nc  &  Ia  & 13.84586 &  -0.33626  &  0.1459  &  0.1231  &  \nodata  &  0.3384  &  15  &  $20.585 \pm 0.171$  &  $18.618 \pm 0.018$  &  $1.967 $  &  $-20.455 $ \\
16467  &  \nodata  &  Ia-photo+z  & 328.59792 &  +0.08437  &  0.2106  &  0.2188  &  \nodata  &  0.3120  &  21  &  $22.307 \pm 0.544$  &  $18.673 \pm 0.013$  &  $3.634 $  &  $-21.918 $ \\
\hline
17435  &  2007ka  &  Ia  & 20.36507 &  +0.02192  &  0.2160  &  0.2210  &  \nodata  &  0.3713  &  18  &  \nodata  &  \nodata  &  \nodata  &  \nodata  \\
17568  &  2007kb  &  Ia  & 313.04022 &  +0.33160  &  0.1486  &  0.1390  &  \nodata  &  0.5406  &  23  &  $22.017 \pm 0.424$  &  $19.651 \pm 0.032$  &  $2.366 $  &  $-19.732 $ \\
18047  &  \nodata  &  Ia-photo  & 22.02004 &  -0.66489  &  0.2592  &  0.3614  &  0.0446  &  0.5526  &  16  &  $22.691 \pm 0.734$  &  $19.566 \pm 0.023$  &  $3.125 $  &  $-22.474 $ \\
18047  &  \nodata  &  Ia-photo  & 22.08698 &  -0.70110  &  0.2970  &  0.3614  &  0.0446  &  0.5014  &  15  &  $22.691 \pm 0.734$  &  $19.566 \pm 0.023$  &  $3.125 $  &  $-22.474 $ \\
18362  &  \nodata  &  Ia-photo  & 10.13752 &  -0.23279  &  0.2350  &  0.2197  &  0.0278  &  0.4781  &  16  &  $22.108 \pm 0.346$  &  $19.446 \pm 0.023$  &  $2.662 $  &  $-21.157 $ \\
18362  &  \nodata  &  Ia-photo  & 10.06133 &  -0.12293  &  0.2269  &  0.2197  &  0.0278  &  0.8767  &  28  &  $22.108 \pm 0.346$  &  $19.446 \pm 0.023$  &  $2.662 $  &  $-21.157 $ \\
18375  &  2007lg  &  Ia  & 11.60084 &  +0.00238  &  0.1189  &  0.1169  &  \nodata  &  0.4614  &  23  &  $20.365 \pm 0.081$  &  $18.167 \pm 0.008$  &  $2.198 $  &  $-20.775 $ \\
18767  &  \nodata  &  Ia-photo  & 4.55214 &  +0.80302  &  0.2377  &  0.2728  &  0.0322  &  0.1752  &  13  &  $22.471 \pm 0.656$  &  $19.118 \pm 0.023$  &  $3.353 $  &  $-22.094 $ \\
18909  &  2007lq  &  Ia?  & 5.79401 &  +0.97633  &  0.2296  &  0.2258  &  \nodata  &  0.1231  &  16  &  $21.935 \pm 0.534$  &  $18.424 \pm 0.013$  &  $3.511 $  &  $-22.255 $ \\
19001  &  \nodata  &  Ia-photo  & 41.65704 &  -0.38006  &  0.2457  &  0.2688  &  0.0243  &  0.7061  &  11  &  \nodata  &  \nodata  &  \nodata  &  \nodata  \\
19341  &  2007nf  &  Ia  & 15.84750 &  +0.31945  &  0.2539  &  0.2339  &  \nodata  &  0.1744  &  22  &  $21.654 \pm 0.232$  &  $18.875 \pm 0.014$  &  $2.779 $  &  $-21.902 $ \\
19969  &  2007pt  &  Ia  & 31.88020 &  -0.24733  &  0.1593  &  0.1744  &  \nodata  &  0.5710  &  15  &  $19.966 \pm 0.067$  &  $18.419 \pm 0.011$  &  $1.547 $  &  $-21.558 $ \\
20111  &  2007pw  &  Ia  & 354.47760 &  +0.22478  &  0.2619  &  0.2468  &  \nodata  &  0.8811  &  37  &  $23.019 \pm 1.046$  &  $19.321 \pm 0.024$  &  $3.698 $  &  $-21.606 $ \\
20232  &  \nodata  &  Ia-photo+z  & 7.08765 &  -0.01711  &  0.2323  &  0.2154  &  \nodata  &  0.3849  &  21  &  $21.814 \pm 0.369$  &  $18.592 \pm 0.012$  &  $3.222 $  &  $-21.956 $ \\
20882  &  \nodata  &  Ia-photo  & 16.97586 &  +0.51513  &  0.2808  &  0.3181  &  0.0186  &  0.5792  &  19  &  $24.582 \pm 1.345$  &  $20.573 \pm 0.042$  &  $4.009 $  &  $-21.086 $ \\
\enddata
\\[-5ex]
\tablecomments{SDSS Id denotes internal candidate designation. 
The horizontal lines delineate SNe discovered during the 3 distinct observing seasons of the \sns.
}
\tablenotetext{a}{``Ia'' and ``Ia?'' refer 
to spectroscopically identified SNe; 
see \citet{Zheng_08}.
``Ia-photo+z'' and ``Ia-photo'' 
refer to photometrically 
identified SNe {\it with} and {\it without}
a spectroscopically measured host galaxy redshift, respectively; 
see \citet{Dilday_10}.}

\tablenotetext{b}{Value has been k-corrected to the rest frame of the galaxy; see \S \ref{sec:kcors}}
\tablenotetext{c}{Errors are shown for photometric SNe Ia. 
For spectroscoipcally measured redshifts, the error is negligible and is 
not listed (\S \ref{sec:sdsssnobs})}
\tablenotetext{d}{Observer frame magnitudes for the SN host galaxy.}
\end{deluxetable}

\acknowledgments

Funding for the SDSS and SDSS-II has been provided by the Alfred P. Sloan Foundation, the Participating Institutions, the National Science Foundation, the U.S. Department of Energy, the National Aeronautics and Space Administration, the Japanese Monbukagakusho, the Max Planck Society, and the Higher Education Funding Council for England. The SDSS Web Site is http://www.sdss.org/.

The SDSS is managed by the Astrophysical Research Consortium for the Participating Institutions. The Participating Institutions are the American Museum of Natural History, Astrophysical Institute Potsdam, University of Basel, University of Cambridge, Case Western Reserve University, University of Chicago, Drexel University, Fermilab, the Institute for Advanced Study, the Japan Participation Group, Johns Hopkins University, the Joint Institute for Nuclear Astrophysics, the Kavli Institute for Particle Astrophysics and Cosmology, the Korean Scientist Group, the Chinese Academy of Sciences (LAMOST), Los Alamos National Laboratory, the Max-Planck-Institute for Astronomy (MPIA), the Max-Planck-Institute for Astrophysics (MPA), New Mexico State University, Ohio State University, University of Pittsburgh, University of Portsmouth, Princeton University, the United States Naval Observatory, and the University of Washington.

This work is based in part on observations made at the 
following telescopes.
The Hobby-Eberly Telescope (HET) is a joint project of the University of Texas
at Austin,
the Pennsylvania State University,  Stanford University,
Ludwig-Maximillians-Universit\"at M\"unchen, and Georg-August-Universit\"at
G\"ottingen.  The HET is named in honor of its principal benefactors,
William P. Hobby and Robert E. Eberly.  The Marcario Low-Resolution
Spectrograph is named for Mike Marcario of High Lonesome Optics, who
fabricated several optical elements 
for the instrument but died before its completion;
it is a joint project of the Hobby-Eberly Telescope partnership and the
Instituto de Astronom\'{\i}a de la Universidad Nacional Aut\'onoma de M\'exico.
The Apache 
Point Observatory 3.5 m telescope is owned and operated by 
the Astrophysical Research Consortium. We thank the observatory 
director, Suzanne Hawley, and site manager, Bruce Gillespie, for 
their support of this project.
The Subaru Telescope is operated by the National 
Astronomical Observatory of Japan. 
The William Herschel 
Telescope is operated by the 
Isaac Newton Group, 
on the island of La Palma
in the Spanish Observatorio del Roque 
de los Muchachos of the Instituto de Astrofisica de 
Canarias. 
Based on observations made with ESO Telescopes at the La Silla or Paranal Observatories under programme IDs
77.A-0437, 
78.A-0325,
79.A-0715, 
and
80.A-0024.
Based on observations made with the Nordic Optical Telescope, operated
on the island of La Palma jointly by Denmark, Finland, Iceland,
Norway, and Sweden, in the Spanish Observatorio del Roque de los
Muchachos of the Instituto de Astrofisica de Canarias.  
Kitt Peak National Observatory, National Optical 
Astronomy Observatory, is operated by the Association of 
Universities for Research in Astronomy, Inc. (AURA) under 
cooperative agreement with the National Science Foundation. 
Based partially on observations made with the Italian Telescopio  
Nazionale Galileo (TNG) operated on the island of La Palma by the  
Fundaci\'on Galileo Galilei of the INAF (Istituto Nazionale di  
Astrofisica) at the Spanish Observatorio del Roque de los Muchachos of  
the Instituto de Astrof\'{\i}sica de Canarias.

This work was supported in part by the Kavli Institute for Cosmological Physics at the University of Chicago through grants NSF PHY-0114422 and NSF PHY-0551142 and an endowment from the Kavli Foundation and its founder Fred Kavli.
This work was also partially supported by the US Department of Energy through grant DE-FG02-08ER41562 to Rutgers University (PI: SWJ).

We thank Chris Miller for making the extended version of the C4 cluster catalog available.
We thank Ben Koester for providing a catalog of cluster member galaxies for the maxBCG clusters.
BD thanks B.~Koester and S.~Hansen for many invaluable discussions.


\end{document}